\documentclass[11pt]{article}
\usepackage{latexsym}
\usepackage{amsmath}
\usepackage{amssymb}
\usepackage{bbm}
\usepackage{graphicx}
\usepackage{subfigure}
\usepackage{dcolumn} 
\usepackage{bm}
\usepackage{cite}
\usepackage{yhmath}
\usepackage{authblk}
\usepackage{mathtools}

\usepackage[usenames,dvipsnames]{color}
\usepackage[]{hyperref}

\usepackage{cleveref}

\usepackage{tikz}
\usepackage[margin=60pt]{geometry}
\usepackage{epsfig, xypic}

\usetikzlibrary{arrows}
\usetikzlibrary{trees}

\usepackage{fancyhdr}
\usepackage{xcolor}
\newsavebox\MBox

\usepackage{setspace}
\newcommand{\be}{\begin{equation}}
\newcommand{\ee}{\end{equation}}
\newcommand{\bea}{\begin{eqnarray}}
\newcommand{\eea}{\end{eqnarray}}
\def \ba {\begin{aligned}}
\def \ea {\end{aligned}}

\def\Tr{{\rm Tr}}

\def\l:{\mathopen{:}\,}
\def\r:{\,\mathclose{:}}

\newcommand{\EQ}[1]{\begin{align}\begin{split} #1 
\end{split}\end{align}}
\newcommand{\eq}[1]{\begin{equation}{ #1
}\end{equation}}

\title{Remarks on 2d Unframed Quiver Gauge Theories}
\author[1]{Peng Zhao\footnote{pzhao@bimsa.cn}}
\author[1,2]{Hao Zou\footnote{hzou@bimsa.cn}}
\affil[1]{Beijing Insitute of Mathematical Sciences and Applications, Beijing 101408, China}
\affil[2]{Yau Mathematical Sciences Center, Tsinghua University, Beijing 100084, China}

\begin{document}

\title{Remarks on 2d Unframed Quiver Gauge Theories}

\maketitle
\abstract{
We study 2d $\mathcal{N}=(2,2)$ quiver gauge theories without flavor nodes. There is a special class of quivers whose gauge group ranks stay positive in any duality frame. We illustrate this with the Abelian Kronecker quiver and the Abelian Markov quiver as the simplest examples. In the geometric phase, they engineer an infinite sequence of projective spaces and hypersurfaces in Calabi-Yau spaces, respectively. We show that the Markov quiver provides an Abelianization of SU(3) SQCD. Turning on the FI parameters and the $\theta$ angles for the Abelian quiver effectively deform SQCD by such parameters. For an Abelian necklace quiver corresponding to SU($k$) SQCD, we find evidence for singular loci supporting non-compact Coulomb branches in the K\"ahler moduli space.}

\newpage

\tableofcontents

\section{Introduction and Summary}

Two-dimensional gauged linear sigma models (GLSMs) with $\mathcal{N} = (2,2)$ supersymmetry \cite{Witten:1993yc} engineer a large class of K\"ahler target geometries. GLSMs with Abelian gauge groups have been well studied in relation to toric geometry and mirror symmetry and then extended to non-Abelian gauge groups. 
Quiver GLSMs have recently been studied from various perspectives \cite{Donagi:2007hi, Jockers:2012zr, Bonelli:2013mma,  Benini:2014mia, Gomis:2014eya, Franco:2015tna, Closset:2017yte, Guo:2018iyr, Closset:2018axq, Guo:2021dlz, Galakhov:2022uyu},  leading to new mathematical structures and physical insights that would not be attainable by considering only a single gauge node.

Infrared dualities play a central r\^ole in relating different quiver GLSMs. In \cite{Benini:2014mia}, building on earlier works on a single gauge node \cite{Hanany:1997vm, Hori:2006dk, Benini:2012ui} and similar dualities in higher dimensions \cite{Seiberg:1994pq, Berenstein:2002fi, Benini:2011mf, Closset:2012eq, Xie:2013lya}, it was found that under a Seiberg-like duality, gauge theories with unitary groups realize a cluster algebra structure. The gauge group ranks and the complexified Fayet-Iliopoulos (FI) parameters transform as cluster variables. The precise matching of parameters has been tested by the sphere partition function \cite{Doroud:2012xw, Benini:2012ui}. Of particular interest is the FI parameter, which parametrizes coordinates on the K\"ahler moduli space.

The study of quiver GLSMs ties closely with the mathematics of quiver varieties. The sphere partition function and the vortex partition function are the generating functions of Gromov-Witten invariants in genus zero \cite{Jockers:2012dk, Bonelli:2013mma}. Dualities imply the equality of such generating functions up to a cluster transformation and came to be known as the Mutation Conjecture in the mathematical literature \cite{Ruan_2017}. For $A_n$ linear quivers, the conjecture has recently been proven in \cite{Zhang:2021hdo}.

Infrared dualities motivate us to ask global questions about quiver GLSMs. There is a special class of quivers that do not break supersymmetry \cite{Xie:2013lya, Benini:2014mia}. In the first part of this note, we introduce the notion of positive GLSM quivers, whose gauge group ranks stay positive in any duality frame. For unframed quivers without a flavor node, the positivity condition is very restrictive. We illustrate this with several examples, which have also been studied extensively in quiver quantum mechanics. For unframed quivers, there is an overall U(1) that decouples in the infrared. The theory lives on a codimension-1 subspace of the multi-dimensional K\"ahler moduli space. 

We shall identify the simplest positive unframed quiver: the Abelian Kronecker quiver, which is also the simplest example in quiver quantum mechanics \cite{Douglas:2000qw, Denef:2002ru}. While the $n$-Kronecker quiver admits an equivalent framed-quiver description as a U(1) theory with $n$ chiral multiplets, the unframed quiver allows an infinite sequence of dualities, reminiscent of a duality cascade in 4d $\mathcal{N}=1$ theories \cite{Klebanov:2000hb}. Unlike in four dimensions where asymptotic freedom makes an essential distinction between Abelian and non-Abelian theories, in two dimensions an Abelian theory can be mapped to a non-Abelian theory by a duality \cite{Hori:2006dk, Jia:2014ffa}. This leads to infinitely-many equivalent descriptions of the projective space. We show that the  K\"ahler cones asymptote to a limiting ray.

The next simplest example is the Abelian Markov quiver, which is a conformal quiver that engineers a Calabi-Yau target space. The 3-Markov quiver has also been studied for D-branes at an orbifold singularity \cite{Douglas:1997de, Diaconescu:1999dt, Douglas:2000qw} and 4d $\mathcal{N}=1$ theories engineered from D-branes wrapping cycles of local Calabi-Yau threefolds \cite{Cachazo:2001sg}, as well as the related quiver quantum mechanics\cite{Douglas:2000qw, Denef:2002ru}.
With a superpotential turned off, we describe the target space geometry at each phase of the K\"ahler parameters, and the flop transitions between them. When a non-degenerate superpotential is introduced, the theory admits an infinite number of equivalent descriptions of a hypersurface in the Calabi-Yau space. The infinite number of duality frames has an interesting global structure of a Bruhat-Tits tree where two theories are related by a path of dualities. The limiting K\"ahler cone can be thought of as the moduli space of the ``boundary" theory.

In the second part of the note, we study another relation between Abelian quivers and non-Abelian theories: Abelianization. 
We observe a coincidence between the vacuum equations of the Markov quiver and SU($3$) SQCD, which is then generalized to the necklace quiver and SU($k$) SQCD. This is akin to various Abelianization methods where the $k$ U(1) gauge nodes can be regarded as the Cartan elements of the SU($k$) theory \cite{Hori:2000kt, Halverson:2013eua, Lee:2013yka}. SU($k$) SQCD with $n$ massless chiral multiplets are known to be singular for certain choices of $k$ and $n$, and supports multiple non-compact Coulomb branches \cite{Hori:2006dk}. We find the same phenomena in the Abelian quiver, and the SQCD result can be recovered by imposing Weyl symmetry and eliminating degenerate solutions.

The Abelian quiver provides more, for one can turn on an FI parameter and a $\theta$ angle for each U(1). One can regard moving into the K\"ahler moduli space as deforming SQCD by an effective complexified FI parameter. As the parameters are tuned, we will find singular points associated with discrete values of the $\theta$ angle. In addition to isolated singularities, there can be continuous families of singular loci. We interpret them as the Lagrange multipliers for unbroken subgroups of SQCD. In higher dimensions, we find multiple disconnected singular loci of mixed dimensions. 

This note is organized as follows. In Section \ref{GLSM}, we introduce quiver GLSMs and review their cluster algebra structure under dualities. We define the notion of a positive GLSM quiver. In Section \ref{examples}, we study the Kronecker quiver and the Markov quiver, and their infinite sequences of duality frames. In Section \ref{necklace}, we present a correspondence between an Abelian quiver and SQCD. We perform a Coulomb branch analysis and compute the number of non-compact Coulomb branches. We deform the Abelian quiver by turning on the complexified FI parameters and find evidence for higher-dimensional singular loci in the K\"ahler moduli space.

\section{Quiver GLSMs and Duality}
\label{GLSM}

\subsection{Quiver GLSMs and Symplectic Quotient}
A quiver is a directed graph consisting of a set of vertices connected by arrows. There are two types of vertices known as gauge nodes and flavor nodes in the context of gauge theory. Each gauge or flavor node corresponds to a U($N_i$) gauge or flavor group factor, respectively. A quiver with flavor nodes is called a framed quiver and one without flavor nodes is called an unframed quiver. We will be interested in quivers that admit a description in terms of cluster algebras \cite{FominZ1}. That is, it is free of any 1-cycles or 2-cycles and can be represented by a signed adjacency matrix $b_{ij}$. One can also consider a more general class of quivers by allowing for 1-cycles and 2-cycles. For example, if we consider a double quiver by including arrows in the opposite direction and adding a 1-cycle for every gauge node, then the result is a Nakajima quiver of an $\mathcal{N}=(4,4)$ gauge theory.

Consider a quiver with $L$ gauge nodes and $M$ flavor nodes. To each arrow $i \to j$ we associate a chiral multiplet $\phi^{i,j}_\alpha$, $\alpha \in \{1, 2, \ldots, [b_{ij}]_+\}$ in the bi-fundamental representation of $\text{U}(N_i) \times \text{U}(N_j)$. Here $[b_{ij}]_+ = \max(b_{ij},0)$ is the number of arrows from $i$ to $j$. With respect to a given gauge node $i$, there are $N^f_i = \sum_{j} [b_{ij}]_+ N_j$ chiral multiplets in the fundamental representation of U($N_i$),  represented by the outgoing arrows, and $N^a_i = \sum_{j} N_j[b_{ji}]_+$  chiral multiplets in the anti-fundamental representation of U($N_i$), represented by the incoming arrows. The absence of gauge anomaly in two dimensions implies that $N^f_i$ can differ from $N^a_i$ and a freedom in choosing the gauge group ranks. When both $i$ and $j$ are flavor nodes, there are $N_i N_j$ gauge-singlet fields for each arrow. We may turn on an FI parameter $\xi_i$ and a $\theta$ angle for each gauge node. They combine into a complexified FI parameter 
\eq{
t_i = 2\pi \xi_i+ i\theta_i \,.
}
The $\theta$ angle induces a background electric field. It is $2\pi$-periodic because pair creation of charged particles will screen the electric field \cite{Coleman:1976uz}.

The supersymmetric ground states are determined by the $D$-term, the $F$-term, and the twisted $F$-term equations. The classical Higgs branch is determined by the vanishing of the $D$-term equations. For the $i$-th gauge node,
\eq{
D_i = -e_i^2\left[\sum_{j \to i} \sum_{\alpha=1}^{b_{ji}} \big(\phi^{j,i}_\alpha \big)^\dagger \big(\phi^{j,i}_\alpha\big)  - \sum_{i \to j^{\prime}} \sum_{\alpha=1}^{b_{ij^{\prime}}} \big(\phi^{i,j^{\prime}}_\alpha \big) \big(\phi^{i,j^{\prime}}_\alpha\big)^\dagger - \xi_i \,\mathbb{I}_{N_i} \right]\,.
\label{$D$-term}
}
Here $e_i$ is the coupling constant.

In the infinite coupling $e_i \to \infty$ limit, the massive modes are integrated out and it is believed that the model flows to a non-linear sigma model with the classical Higgs branch of the GLSM \cite{Witten:1993yc} as its target space. The FI parameter controls the size of the target space and is the K\"ahler class of the sigma model. It runs logarithmically, with a one-loop exact $\beta$ function proportional to
\eq{
\beta_i \propto N^a_i - N^f_i \,.
}
When all the $\beta$ functions vanish, the theory flows to a superconformal fixed point in the infrared, with a Calabi-Yau target space.

The classical Higgs branch is a quiver variety that arises by a symplectic quotient construction. Each node corresponds to a $\mathbb{C}^{N_i}$ and each arrow is a map $\text{Mat}(\mathbb{C}^{N_i}, \mathbb{C}^{N_j})$. Let 
\eq{V = \bigoplus_{\text{all arrows } i \to j} \mathbb{C}^{N_i \times N_j}, \qquad G = {\rm U}(N_1)\times {\rm U}(N_2)\times \cdots \times {\rm U}(N_L) \,.}
The gauge symmetry induces a momentum map $\mu(\phi): V \to \mathfrak{g}^*$, defined by the $D$-term equations \eqref{$D$-term}. This defines a quiver variety as the symplectic quotient of the level set of the momentum map by the gauge symmetry
\eq{
\mu^{-1}(\vec \xi) / G \,.
}

A standard example is the flag variety. It is the configuration space of an embedding sequence $\mathbb{C}^{N_1} \subset \mathbb{C}^{N_2} \subset \cdots \subset \mathbb{C}^{N_L} \subset \mathbb{C}^N$, which is called a flag. It is naturally realized as the geometric phase of a linear quiver with a flavor node at the end \cite{Donagi:2007hi}.

When the quiver is unframed, $b_{ij}$ is an $L \times L$ skew-symmetric matrix. The FI parameters are not all independent but satisfy $\sum_{i=1}^L N_i\, \xi_i = 0$. By taking a matrix trace, we see that $\text{Tr}_{N_i} \left(\big(\phi^{i,j}_\alpha \big) \big(\phi^{i,j}_\alpha\big)^\dagger\right)$ with $b_{ij} > 0$ appears with a minus sign in the equation for $\xi_i$, and with a plus sign in the equation for $\xi_j$. An overall U(1) decouples and the gauge symmetry is ${\rm S}\big[{\rm U}(N_1)\times {\rm U}(N_2)\times \cdots \times {\rm U}(N_L)\big]$.

There is a U(1) R-symmery that act on the chiral fields by $\phi^{i,j}_\alpha \to \lambda^{r^{i,j}_\alpha} \phi^{i,j}_\alpha$, where $\lambda \in \text{U}(1)$. The R-charges $r^{i,j}_\alpha \in \mathbb{R}$ assign gradings to the arrows. We may turn on a quasi-homogeneous superpotential $W(\phi)$ consistent with the R-charge assignments of the fields such that every term has R-charge 2. 
The $F$-term equations are the critical loci of the superpotential 
\eq{
\frac{\partial W}{\partial \phi^{i,j}_{\alpha}} = 0 \,,
}
and define a hypersurface in the quiver variety.

The theory has a classical Coulomb branch of dimension $\sum^L_{i=1} N_i$ parametrized by the complex scalars $\sigma_i$ each taking values in the Cartan subalgebra of the vector multiplet of U($N_i$). For an unframed quiver, the overall U(1) decouples and the classical Coulomb branch has dimension $\sum^L_{i=1} N_i - 1$. In the quantum theory, the Coulomb branch is generically lifted and the supersymmetric vacua are discrete, although non-compact Coulomb branches may survive. It is governed by an effective twisted superpotential that can be computed from the one-loop renormalization of the tree-level FI term $\widetilde W = -\sum_i t_i\, \sigma_i$: 
\EQ{\widetilde W_\text{eff} = 
&-\sum_{\text{all gauge nodes } i}\left[ t_i \, \sum_{a =1}^{N_i} {(\sigma_i)}_a - i\pi \sum_{a < b}^{N_i}\big({(\sigma_i)}_a - {(\sigma_i)}_b \big) \right] \\ 
&- \sum_{\text{all arrows } i \to j }\sum_{a=1}^{N_i} \sum_{b=1}^{N_j} \big( (\sigma_i)_a - (\sigma_j)_b \big) \Big[\log \big(-i\big( (\sigma_i)_a - (\sigma_j)_b \big)\big) -1 \Big] \,.
\label{twistedW}
}
The second term in the first line is the contribution from the off-diagonal vector multiplets that contributes an effective $\theta$ angle $(N_i+1)\pi$. When $j$ is a flavor node, then $\sigma_j$ is the twisted mass. The $-i$ factor in the logarithm matches with the convention of the sphere partition function in \cite{Benini:2012ui}.

The Coulomb vacua are given by the solutions of the twisted $F$-term equations 
\eq{
\exp\left(\frac{\partial \widetilde W_\text{eff}}{\partial (\sigma_i)_a}\right) = 1 \,,
\label{twistedF}
}
modulo action by the Weyl group. Because the effective twisted superpotential is derived based on the assumptions that the gauge group is broken to its Cartan subgroup and all massive fields can be integrated out, we should exclude the solutions $(\sigma_i)_a = (\sigma_i)_b$ for all $a \ne b$ and $(\sigma_i)_a = (\sigma_j)_a$ for all $a$ between neighboring nodes with $b_{ij} \ne 0$. 

\subsection{Duality as Cluster Transformation}

We summarize the main features of the Seiberg-like duality for 2d quiver gauge theories below, focusing on the cluster algebraic structure. 
Dualities for general quivers with 1-cycles, i.e. adjoint chiral multiplets, have also been studied \cite{Benini:2014mia, Gomis:2014eya}, although they do not realize a cluster algebra. See \cite{Benini:2016qnm, Park:2016dpb} for comprehensive reviews, including localization computations essential to precisely match the parameters.

\begin{figure}[!h]
\centering
\includegraphics[scale=0.4]{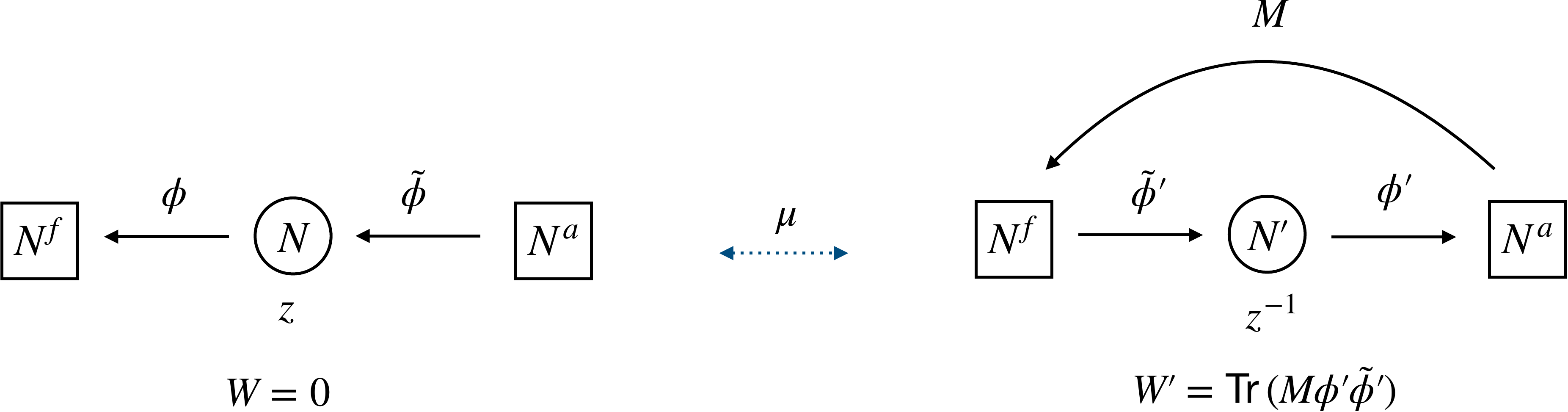}
\caption{The Seiberg-like duality for a U($N$) gauge theory with $N^f$ fundamental and $N^a$ anti-fundamental chiral multiplets. The quiver undergoes a mutation, generating $N^f N^a$ gauge-singlet fields and a superpotential. The gauge group ranks and the FI parameters transform as cluster variables.}
\label{Seiberg}
\end{figure}

A Seiberg-like duality generally maps a gauge theory $\mathcal{A}$ with $N^f$ chiral fields $\phi$ in the fundamental representation of the gauge group and $N^a$ chiral fields $\tilde \phi$ in the anti-fundamental representation to another gauge theory $\mathcal{B}$ with $N^f$ anti-fundamental fields $\tilde \phi'$, $N^a$ fundamental fields $\phi'$, $N^f N^a$ gauge-singlet fields $M$, and a superpotential of the form $W = \Tr (M  \phi' \tilde \phi' )$. The quivers of dual theories are shown in Fig. \ref{Seiberg}, which realize a quiver mutation. A mutation $\mu_k$ is a local operation on a gauge node in three steps:
\begin{enumerate}
\item Reverse the directions of all incoming and outgoing arrows.
\item For each path between $j \to k \to i$, form a cycle by connecting $i$ to $j$. 
\item Remove pairs of opposite arrows to eliminate 2-cycles.
\end{enumerate} 
Flavor nodes do not mutate and are also called frozen nodes. 

\begin{figure}[!h]
\centering
\includegraphics[scale=0.4]{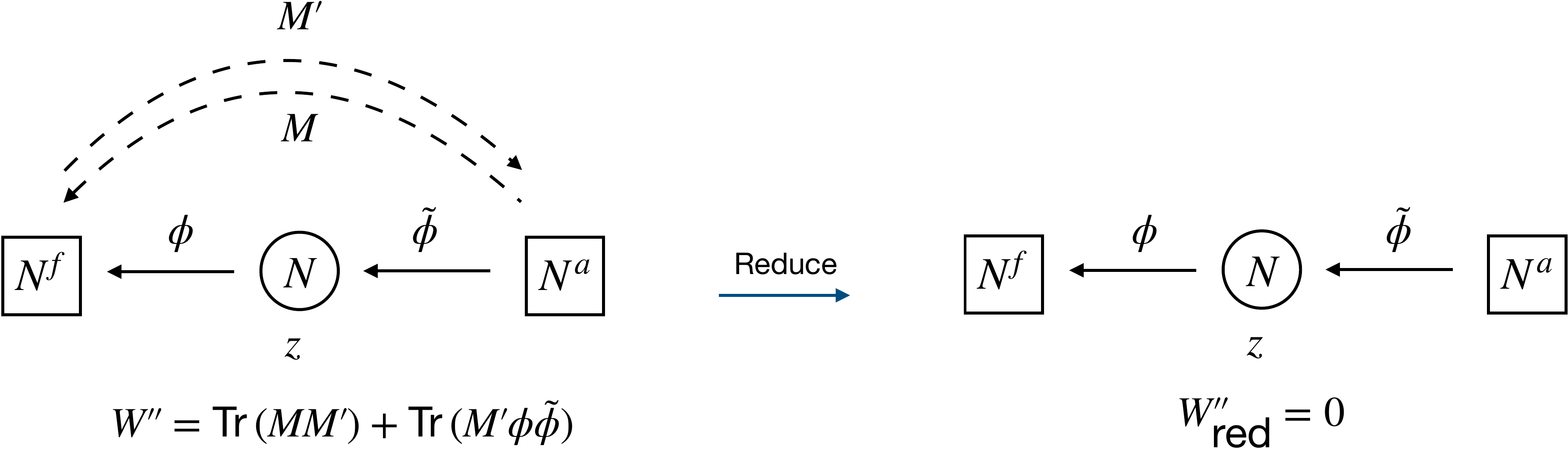}
\caption{Integrating out quadratic terms in the superpotential removes 2-cycles in the quiver. Superpotentials where this can be consistently done at any step are called non-degenerate potentials.}
\label{Seiberg2}
\end{figure}

To incorporate superpotentials consistent with the duality, we consider quivers with non-degenerate superpotentials $(b_{ij}, W)$ \cite{Derksen1}. In Step 1, the fields $\phi^{i,k}_\alpha$ and $\phi^{k,j}_\alpha$ are mapped to dual fields $(\phi^{i,k}_\alpha)' = \phi^{k,i}_\alpha$ and $(\phi^{k,j}_\alpha)' = \phi^{j,k}_\alpha$ of R-charges $1-r^{i,k}_\alpha$ and $1-r^{k,j}_\alpha$, respectively. Starting from the superpotential of theory $\mathcal{A}$, we replace all U($N_k$)-invariant composite operators $\phi^{k,j}_\alpha \phi^{i,k}_\beta$ by the gauge-singlet fields $M^{i,j}_{\alpha \beta}$. 
For each cycle generated in Step 2, we add a new term by tracing over the fields
\eq{
W(\phi^{k,j}_\alpha\phi^{i,k}_\beta, \phi_\text{others}) \to W(M^{i,j}_{\alpha \beta}, \phi_\text{others} ) + \sum_{\alpha, \beta} \text{Tr}(M^{i,j}_{\alpha \beta} \phi^{k,i}_\beta \phi^{j,k}_\alpha )\,.
\label{dualW}
}
The gauge-singlets $M^{i,j}_{\alpha \beta}$ in theory $\mathcal{B}$ have R-charge $r^{k,j}_\alpha + r^{i,k}_\beta  $ so the dual superpotential also has R-charge 2.
In Step 3, quadratic terms in \eqref{dualW} corresponding to 2-cycles in the quiver are integrated out, as shown in Fig. \ref{Seiberg2}. A superpotential is non-degenerate if all 2-cycles can be cancelled in this way.
This is a very non-trivial requirement. Suppose we started from theory $\mathcal{B}$ but without a superpotential $W'=0$, then we will not be able to remove the gauge singlets $M$ to obtain theory $\mathcal{A}$. A generic superpotential would also not be compatible with the duality.

The foregoing description of a mutation on a quiver with a non-degenerate superpotential is common to Seiberg duality and similar dualities in other dimensions. To be specific to 2d $\mathcal{N}=(2,2)$ theories we shall introduce GLSM data. A GLSM quiver with unitary groups is labeled by $(b_{ij}, W, N_i, z_i)$, where $N_i$ and $z_i$ (defined below) label the ranks and the complexified FI parameters of the nodes, respectively (one can formally introduce FI parameters for the flavor nodes). 

Under a duality, the gauge group ranks transform as tropical cluster variables \cite{FockGoncharov, Xie:2013lya}
\eq{N_i' = \max(N^f_i, N^a_i) - N_i \,.
\label{rank}
}
The complexified FI parameters are the coordinates on the K\"ahler moduli space
\eq{
z_i = (-1)^{N^f_i-N_i} e^{-t_i} \,.
}
The extra minus signs in the definition ensure that the K\"ahler coordinates transform as dual cluster variables 
\eq{
z_i' = 
\begin{cases}
z_k^{-1} \qquad &  \text{if } i = k \\
z_i z_k^{[-b_{ki}]_+}\qquad & \text{if } i \ne k \text{ and } N^a_k > N^f_k  \\
z_i z_k^{[b_{ki}]_+} (1+z_k)^{-b_{ki}} \qquad & \text{if } i \ne k \text{ and } N^a_k = N^f_k \\
z_i z_k^{[b_{ki}]_+}  \qquad & \text{if } i \ne k \text{ and } N^a_k < N^f_k
\label{FI}
\end{cases} 
\,.
}
Because the cluster transformation preserves the Poisson bracket $\{z_i, z_j\}:= b_{ij}z_i z_j$, this implies a cluster algebra structure on the K\"ahler moduli space where mutations act as canonical transformations. 

The proposal has been tested by sphere partition functions computed exactly using supersymmetric localization \cite{Doroud:2012xw, Benini:2012ui}. The partition functions factorize into sums of products of vortex partition functions. The vortex partition function identities imply that the sphere partition functions agree, up to a non-trivial contact term. The contact terms imply that neighboring FI parameters transform according to \eqref{FI}.

Of course, to fully establish a physical duality, one still needs to match unprotected quantities such as the K\"ahler potential (See \cite{Aharony:2016jki} for examples). The matching of the partition functions across dualities and the cluster transformation of variables provide highly non-trivial evidence for the duality. Mathematically, it has been given a precise formulation that the generating functions of the Gromov-Witten theory on the quiver varieties are equivalent up to a cluster transformation of the parameters \cite{Ruan_2017}. The vortex partition function is precisely the quasimap $I$-function in genus zero. For $A_n$ linear quivers corresponding to flag varieties, the conjecture has recently been proven in \cite{Zhang:2021hdo}.

We remark that for conformal quivers where $N^f_k = N^a_k$, the point $z_k = -1$ indicates a singularity where the contact terms diverge and dualities may break down. This can happen when $t_k = 0$ and $N^f_k - N_k$ is an odd integer, or when $t_k = i\pi$ and $N^f_k-N_k$ is an even integer. 

The twisted chiral ring is generated by gauge-invariant operators $\Tr(\sigma_i^n)$, $n=1, 2, \ldots, N_i$, encoded in the Baxter-Chern polynomial
\eq{
Q_i(x) = \det (x-\sigma_{i})\,.
}
Geometrically, they are the Chern classes for the vector bundles over the target manifold that generate the cohomology ring.
For a quiver, one can attach a $Q_i(x)$ polynomial to each node. Under the duality map, $Q_i$ also mutates similarly as cluster variables 
\eq{
C_k(z_k) Q_k(x) Q_k'(x)  =\prod_{k \to j} Q_j(x)^{b_{kj}} + i^{N^a_k - N^f_k} z_k\prod_{j' \to k} Q_{j'}(x)^{b_{j'k}} \,.
\label{Baxter}
}
Here $C_k(z)$ can be read off from the coefficient of the highest-degree term on the RHS.

One can make the Baxter-Chern polynomials transform exactly as cluster variables by absorbing the K\"ahler moduli $z_i$ into the definition. The operator maps between dual theories can be read off by comparing the polynomial coefficients. 
Evaluating \eqref{Baxter} at the roots of $Q_k$ leads to the vacuum equations of the theory \eqref{twistedF}.

It is an important question which roots of $Q_k$ correspond to the vacua of the theory. In general, there are more solutions than those counted by the Witten index. It is generally assumed that only distinct solutions should be counted and that they account for all the vacua. This assumption, however, has only been tested in specific examples. For U($k$) SQCD with $n$ chiral multiplets, the Witten index $n \choose k$ is the Euler characteristic of the Grassmannian $G(k,n)$. A more intricate argument for SU($k$) SQCD is given in \cite{Hori:2006dk}. For U($N$) SQCD with $(N^f, N^a)$ fundamental and anti-fundamental chiral multiplets, the Witten index was computed to be $N^f \choose N$ when $N^f > N^a$ \cite{Benini:2013xpa}. It follows by charge conjugation that the index should be $\max(N^f, N^a) \choose N$ in the general case. Since the discrete Coulomb vacua computed in the weakly-coupled region account for all the supersymmetric ground states, we conclude in this case that we have found all the supersymmetric vacua.

While we cannot preclude the possibility of degenerate vacua in the general case, we will proceed with this assumption. 
  Because the RHS is a polynomial of degree $\max(N^f_k, N^a_k)$ and one needs to choose $N_k$ distinct roots, there are $\max(N^f_k, N^a_k) \choose N_k$ vacua.

\subsection{Positive GLSM Quivers}

\begin{figure}[!h]
\centering
\includegraphics[scale=0.4]{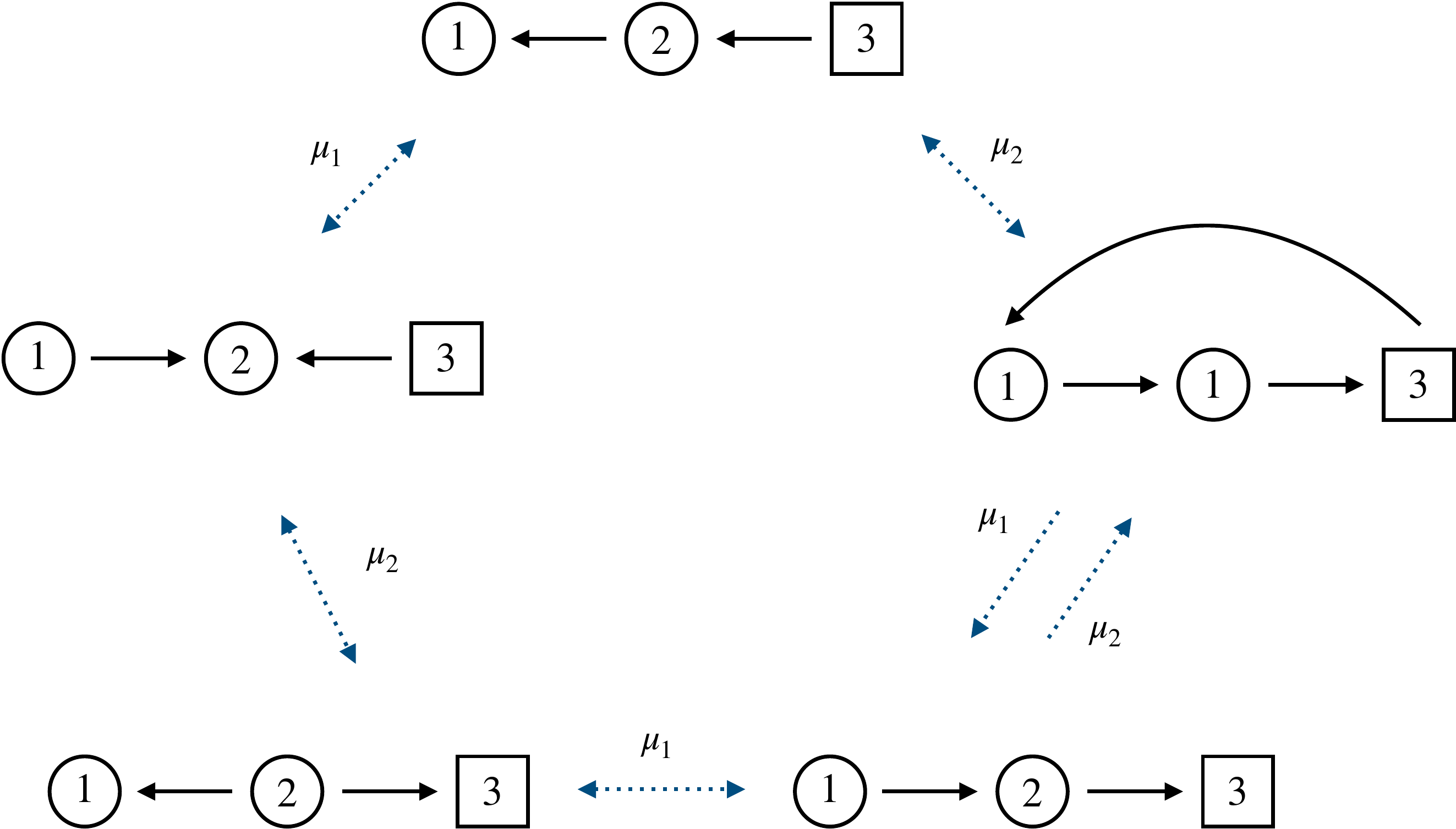}
\caption{All the five duality frames of the flag variety $F(1,2,3)$.}
\label{flag}
\end{figure}

An interesting question concerns classification. When 
\eq{\max(N^f_k, N^a_k) < N_k\,,}
i.e. $N'_k < 0$, there is no supersymmetric ground state and supersymmetry is broken \cite{Xie:2013lya, Benini:2014mia}. 
This naturally leads us to introduce the notion of {\bf positive GLSM quivers}. A quiver defining a GLSM is said to be positive if all the gauge group ranks stay positive in any duality frame. 
Flag varieties provide examples of positive GLSM quivers. Dualities for linear quivers realizing flag manifolds have been studied in \cite{Donagi:2007hi, Guo:2018iyr, Zhang:2021hdo}. An example involving five duality frames is shown in Fig. \ref{flag}.  Two of the dual frames were analyzed in \cite{Ohmori:2018qza}. Note that while it is typically defined by a linear quiver involving non-Abelian gauge groups, there is an Abelian quiver description. This is an example of an Abelian/non-Abelian duality. There are five duality frames in total because the gauge nodes define an $A_2$ cluster algebra, and the number of duality frames of a linear quiver is the Catalan number \cite{Fomin_2003}.  

The classification of finite-type cluster algebras \cite{Fomin_2003} immediately leads to the conclusion that a quiver has finitely-many duality frames if the underlying graph of the gauge nodes are of finite Dynkin type. But this is not a necessary condition. As we shall see for unframed quivers, $D$-term equations can constrain the FI parameters such that the Abelian Kronecker quiver is self-dual, although the quiver is of affine type. The Abelian Markov quiver has an infinite number of duality frames with different FI parameters, but only a single quiver in its mutation class.

\begin{figure}[!h]
\centering
\includegraphics[scale=0.4]{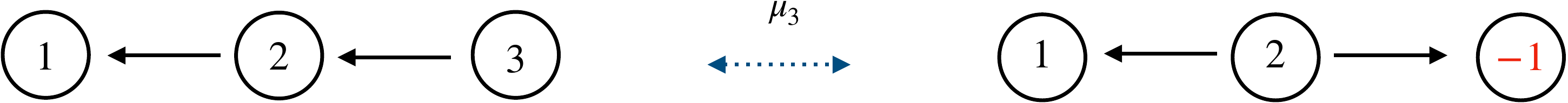}
\caption{If the flavor node in the flag variety were gauged, then dualizing the node would turn the gauge group rank negative. This quiver is not a positive GLSM quiver.}
\label{flag2}
\end{figure}

For unframed quivers, positivity is a very strong condition. For $A_n$ linear quivers without a flavor node, we always get $N_i' \le 0$ in some duality frame, as shown in Fig. \ref{flag2}. Informally, we need enough ``flavors" for each gauge node to ensure that the gauge group rank stays positive after a tropical cluster transformation. This motivates us to study quivers with multiple arrows. 
While we do not have a solution to the combinatorial problem of classifying positive GLSM quivers, we will study several examples, which already exhibit rich properties.

\section{Positive Unframed Quivers and Infinite Duality Chains} 
\label{examples}

\subsection{The Kronecker Quiver}
\begin{figure}[!h]
\centering
\includegraphics[scale=0.5]{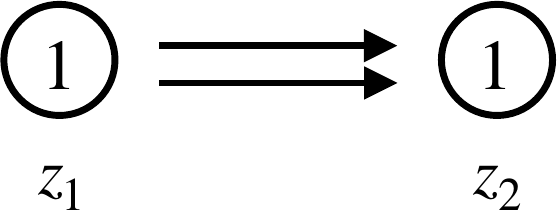}
    \caption{The Kronecker quiver with an Abelian gauge group.}
      \label{kroneckercp1}
\end{figure}

Let us consider the Kronecker quiver, also known as the Dynkin quiver of affine $A_1$ type, as shown in Fig. \ref{kroneckercp1}.
It has been extensively studied in quiver representation theory and quiver quantum mechanics, which is the dimensional reduction of the 2d theory. When the gauge nodes are Abelian, both the quiver and the gauge group rank map to itself as \eqref{rank}, up to an overall $\mathbb{Z}_2$ symmetry that exchanges the nodes. Thus the Abelian Kronecker quiver is the \emph{simplest} positive unframed quiver. The D-term equations coincide with that of the quiver quantum mechanics \cite{Denef:2002ru}, and we will review the Higgs branch analysis below.

It is a $\text{U}(1) \times \text{U}(1)$ gauge theory with two bi-fundamental chiral multiplets. We use the notation $\phi_\alpha := \phi^{1,2}_{\alpha}$ for the bi-fundamental fields.
The corresponding GLSM data is given in Table \ref{data1}
\begin{table}[!h]
\center
       \begin{tabular}{c|cc|c}
        			&  $\phi_1$ & $\phi_2$ & FI \\ \hline
        ${\rm U}(1)_1$ & 1  & 1 &$ \xi_1$ \\
        ${\rm U}(1)_2$ & $-1$  & $-1$  & $\xi_2$  
    \end{tabular}
    \caption{GLSM data for the Abelian Kronecker quiver.}
    \label{data1}
\end{table}

The $D$-term equations for to the two U(1) gauge nodes are
\eq{
     |\phi_1|^2 + |\phi_2|^2 = \xi_1, \qquad -|\phi_1|^2 - |\phi_2|^2 = \xi_2 \,.
}
These two relations are essentially one relation, restricting the theory to the locus $\xi_1 +\xi_2 = 0$. This redundancy implies that there would be a decoupled U(1) in the infrared. One way to see this is to transform to a new basis $Q_\pm = (Q_1 \pm Q_2)/2$ such that $\phi_1$ and $\phi_2$ are no longer charged under $\text{U}(1)_+$. The GLSM data is given in Table \ref{data2}
\begin{table}[!h]
\center
       \begin{tabular}{c|cc|c}
         &  $\phi_1$ & $\phi_2$ & FI \\ \hline
        ${\rm U}(1)_-$ & 1  & 1 &$ \xi_-$ \\
        ${\rm U}(1)_+$ & $0$  & $0$  & $0$  
    \end{tabular}
    \caption{GLSM data for the Abelian Kronecker quiver in the new basis.}
    \label{data2}
\end{table}

\begin{figure}[!h]
\centering
\includegraphics[scale=0.5]{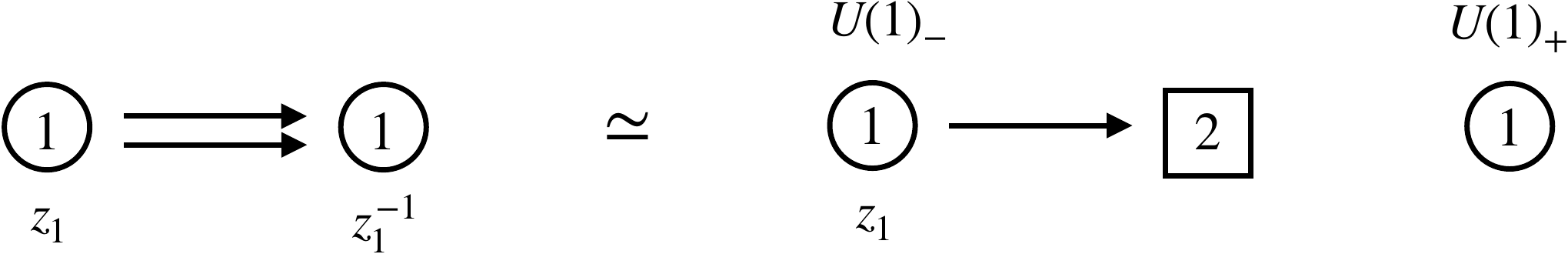}
    \caption{The Abelian Kronecker quiver is equivalent to a framed quiver. An overall U(1) decouples.}
    \label{KroneckerDecouple}
\end{figure}

The FI parameter also combines as $\xi_- := (\xi_1 - \xi_2)/2 = \xi_1$. The basis transformation gives the new $D$-term constraints as
\eq{
|\phi_1|^2 + |\phi_2|^2 = \xi_- \,.
}
In this new basis $\text{U}(1)_+$ is totally decoupled while the $\text{U}(1)_-$ part remains. It is the same as a U(1) theory with 2 chiral multiplets,  The chiral fields can be traded for a flavor node, as shown in Fig. \ref{KroneckerDecouple}.

\begin{figure}[!h]
\centering
\includegraphics[scale=0.5]{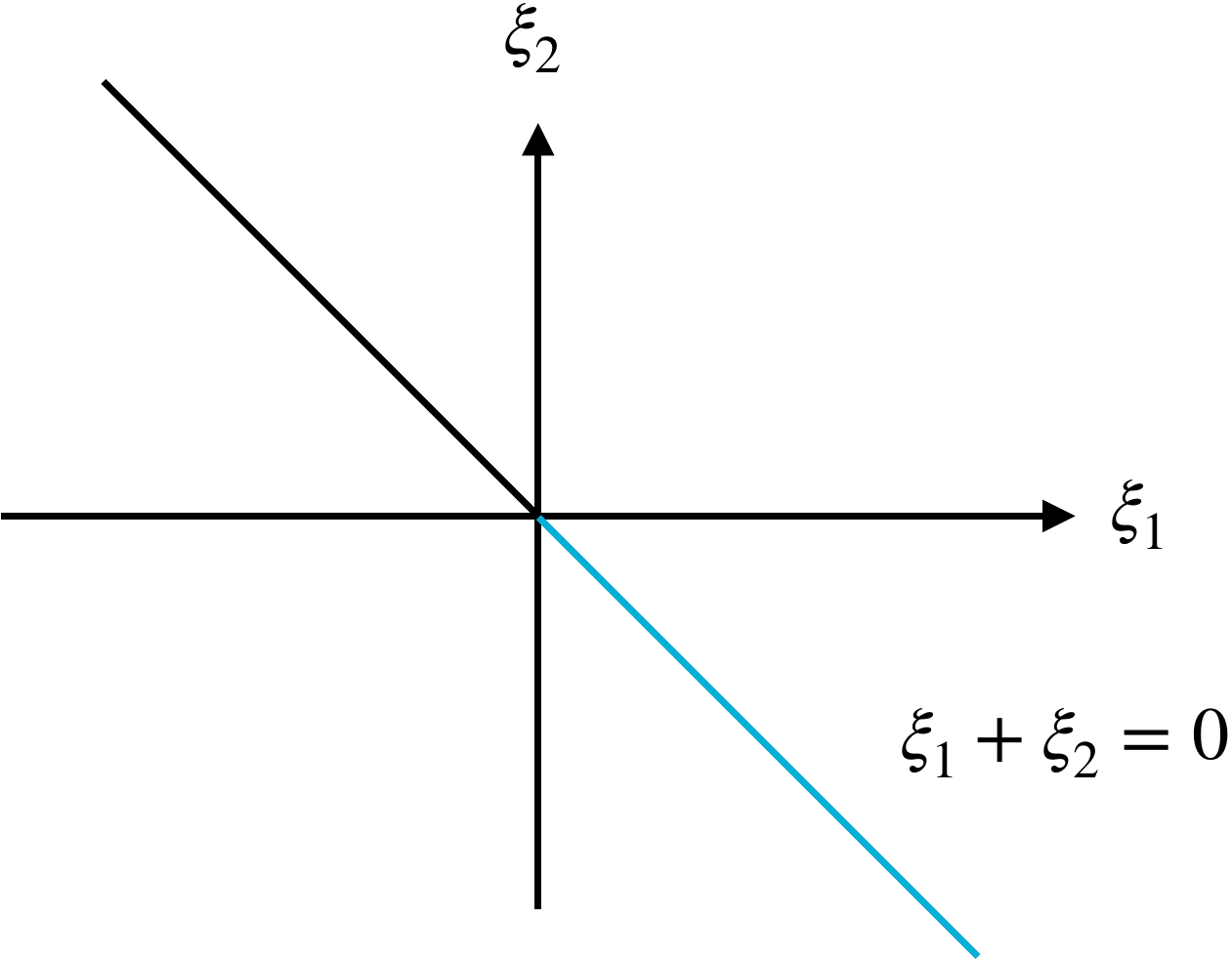}
    \caption{The phase diagram for the Abelian Kronecker quiver. The K\"ahler cone is the ray $\xi_1 > 0, \xi_1 + \xi_2 =0$.}
    \label{KroneckerPhase}
\end{figure}

The phase diagram is shown in Fig. \ref{KroneckerPhase}. Instead of a two-dimensional K\"ahler moduli space, the theory actually lives on a codimension-1 subspace. In the geometric phase $\xi_- \gg 0$, it flows to a $\mathbb{CP}^1$ non-linear sigma model.

The  twisted chiral ring relations \eqref{Baxter} are
\eq{
-z_1 Q_1(x)Q'_1(x) = 1 - z_1 Q^{2}_2(x), \qquad  Q_2(x)Q'_2(x) = Q^{2}_1(x) - z_2 \,.
}
The vacuum equations are obtained by evaluating this at the roots of $Q_1$ and $Q_2$ respectively 
\eq{
    (\sigma_1 - \sigma_2)^2 = z_1^{-1},\qquad
    (\sigma_2 - \sigma_1)^2 = z_2\,.
}
Again, consistency of the equations implies $z_2 = z_1^{-1}$. By shifting to $\sigma_- = \sigma_1 - \sigma_2$, we see that it is equivalent to the quantum cohomology of $\mathbb{CP}^1$. The relation between the Baxter-Chern polynomials is novel and it would be nice to understand its geometric meaning.

Under a duality, the FI parameters map accoording to \eqref{FI}
\eq{z_1 \to z_1^{-1}, \qquad z_2 \to z_2 z_1^2 = z_1\, .} 
Thus the Abelian Kronecker quiver is also the simplest example of a self-dual theory. Instead of having an infinite number of duality frames as one may expect of an affine quiver, the theory maps to itself under the duality.

\begin{figure}[h]
\centering
\includegraphics[scale=0.5]{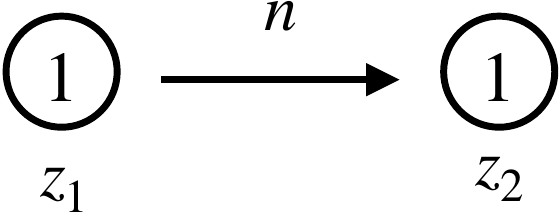}
    \caption{The Abelian $n$-Kronecker quiver.}
    \label{nKronecker}
\end{figure}

\paragraph{The $n$-Kronecker Quiver}
The discussion can be generalized to the $n$-Kronecker quiver containing $n$ arrows between the nodes as shown in Fig. \ref{nKronecker}. 

The $D$-term equations are 
\eq{
\sum_{i=1}^{n} |\phi_i|^2 = \xi_1\,,\qquad -\sum_{i=1}^{n} |\phi_i|^2 = \xi_2 \,.
}
The same analysis shows that the theory is defined on the locus $\xi_1 + \xi_2 = 0$ and is equivalent to a $\text{U}(1)_-$ theory with $n$ chiral multiplets, with a decoupled $\text{U}(1)_+$. The GLSM flows to a $\mathbb{CP}^{n-1}$ sigma model in the infrared. The  phase diagram is the same as in Fig. \ref{KroneckerPhase}.

\begin{figure}[!h]
\centering
\includegraphics[scale=0.5]{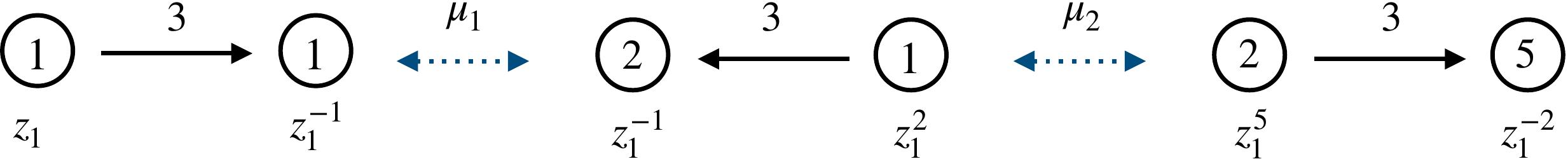}
\caption{The infinite mutation sequence of the Abelian $3$-Kronecker quiver.}
\label{n-Kronecker}
\end{figure}

Here the advantage of the unframed description is that while the framed description can only be dualized once to produce the familiar $\mathbb{CP}^{n-1} \leftrightarrow G(n-1,n)$ duality, the Kronecker quiver can be dualized an infinite number of times by mutating the nodes alternatingly. We obtain an infinite chain of equivalent descriptions of $\mathbb{CP}^{n-1}$, with increasing gauge group ranks but also larger gauge symmetries to divide by. The FI parameters map as  \eqref{FI}. This is shown in Fig.  \ref{n-Kronecker}. 

The mutation of the $n$-Kronecker quiver has also been studied in quiver quantum mechanics \cite{Galakhov:2013oja, Manschot:2013dua, Cordova:2015zra, Kim:2015oxa}. 
While the analysis of the D-term equations is the same and the mutation is very similar in this case, we note that for general quivers the mutation rules in 1d are slightly different \cite{Alim:2011kw, Manschot:2013dua, Kim:2015fba}. The 1d theory does not have a $\theta$ angle, whereas in 2d the complexified FI parameters transform according to \eqref{FI}.

\begin{figure}[!h]
\centering
\includegraphics[scale=0.5]{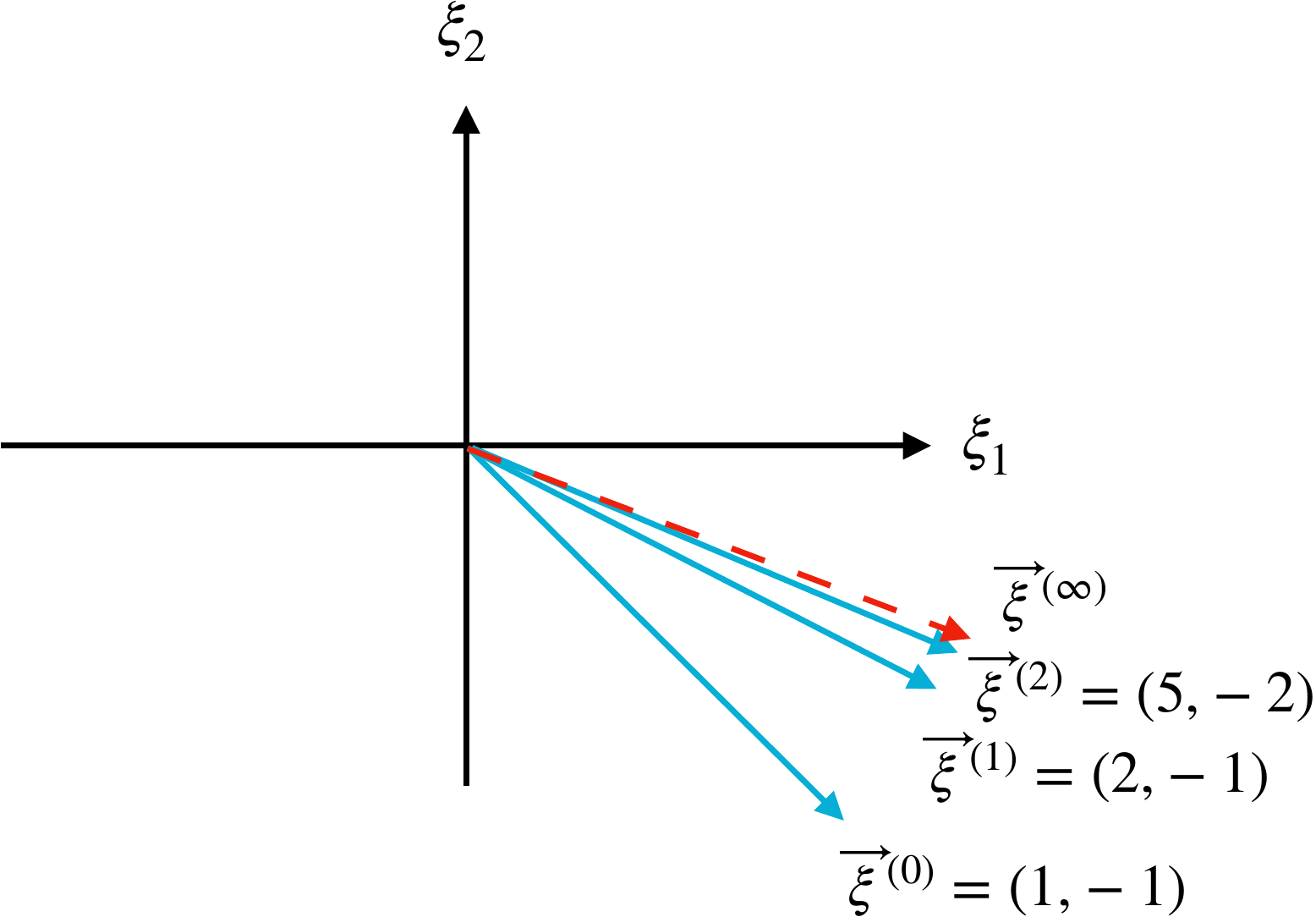}
    \caption{The one-dimensional K\"ahler cone of the Abelian $n$-Kronecker quiver asymptotes to a limiting ray under an infinite sequence of dualities.}
    \label{LimitRay}
\end{figure}

We apply a $\mathbb{Z}_2$ symmetry on every other quiver in this sequence such that the arrows always point from the first to the second node. The next few theories in this chain are a $\text{U}(1) \times \text{U}(n-1)$ theory supported on the locus $\xi_1 + (n-1)\xi_2 = 0$, a $\text{U}(n-1) \times \text{U}(n^2-n-1)$ theory supported on the locus $(n-1)\xi_1 + (n^2-n-1)\xi_2 = 0$, etc.
We remark that the one-dimensional K\"ahler cone defined by $\vec \xi  = (\xi_1, \xi_2)$ with $\xi_1 > 0$ asymptotes to a limiting value.  The FI parameters after the $n$-th mutation $\xi^{(n)}_i$ satisfy the recursion relation 
\eq{
\left(\xi_1^{(n+1)},\, \xi_2^{(n+1)} \right) = \left( n \, \xi_1^{(n)} + \xi_2^{(n)}, \, -\xi_1^{(n)} \right) \,.
}
with $\vec{\xi}^{(0)} = (1, -1)$. Its consecutive ratio $r^{(n)} : = \xi^{(n)}_1/\xi^{(n)}_2$ satisfies the continued-fraction relation
\eq{
r^{(n+1)} = -n - \frac{1}{r^{(n)}} \,.
}
We may solve for the asymptotic consecutive ratio $r^{(\infty)}$ and conclude that the limiting  K\"ahler cone is along the direction $\vec \xi^{(\infty)} = \left( n + \sqrt{n^2-4})/2,-1 \right)$, as shown in Fig. \ref{LimitRay}.

It is interesting that although most of the quivers in the mutation sequence appear non-Abelian, they are mutation-equivalent to an Abelian quiver that defines a toric variety. Such an Abelian/non-Abelian duality is a special feature of 2d theories \cite{Hori:2006dk, Jia:2014ffa}. It is a non-trivial question how to determine whether a non-Abelian quiver is dual to an Abelian quiver, especially when the GLSM cannot be realized as a cluster algebra, such as when the quiver contains 1-cycles or 2-cycles. 

\subsection{The Markov Quiver}

\begin{figure}[!h]
    \centering
    \includegraphics[scale=0.5]{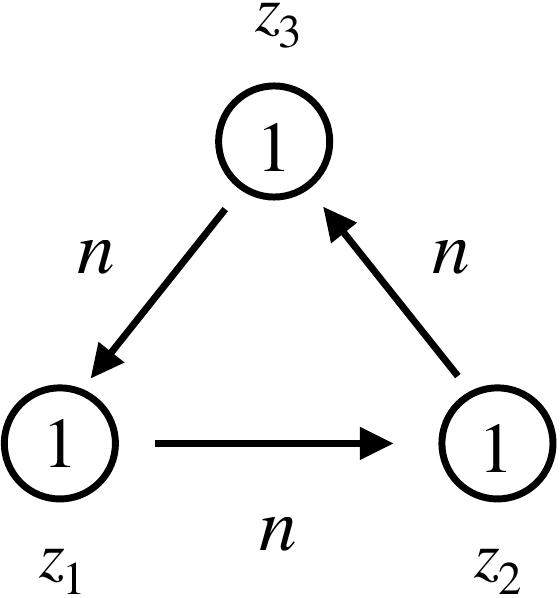}
    \caption{The Abelian $n$-Markov quiver.}
    \label{Markov}
\end{figure}

The simplest positive unframed quiver with three nodes is the cyclic triangle quiver with two arrows between nodes, also known as the Abelian Markov quiver. For the $n$-Markov generalization containing $n$ arrows between the nodes, the number of arrows and gauge group ranks satisfy Markov-type Diophantine equations 
 \EQ{
& b_{12}^2 + b_{23}^2 + b_{31}^2 - b_{12} \, b_{23}\, b_{31} = n^2(3-n)\,, \\
& N_1^2 + N_2^2 + N_3^2 - n\, N_1 N_2 N_3 =3-n \,,
}
 in all duality frames. The 2-Markov quiver is particularly interesting because it also arises from the ideal triangulation of a once-punctured torus \cite{Fomin_2008}, and admits a non-degenerate superpotential \cite{Derksen1}.  It also appears as the BPS quiver of the 4d $\mathcal{N}=2^*$ theory \cite{Alim:2011kw}, whose BPS spectrum is infinite in all chambers and is still not well understood.
 
Let us first turn off the superpotential and study the geometry of the toric variety.
We use $\phi_{1}, \phi_{2}$ to denote $\phi^{1,2}_{\alpha}$, $\phi_{3}, \phi_{4}$ to denote $\phi^{2,3}_{\alpha}$ and $\phi_{5}, \phi_{6}$ to denote $\phi^{3,1}_{\alpha}$. The GLSM data for the Abelian Markov quiver is given in Table \ref{GLSMMarkov}. One can immediately check that this satisfies the Calabi-Yau condition. 
\begin{table}[!ht]
    \centering
    \begin{tabular}{c|cccccc|c}
         &$\phi_1$ & $\phi_2$ &$\phi_3$ & $\phi_4$  &$\phi_5$ & $\phi_6$ & FI\\ \hline
        $\text{U}(1)_1$ & 1  & 1 & 0  & 0 & $-1$ & $-1$ & $\xi_1$\\
        $\text{U}(1)_2$ & $-1$ & $-1$ & 1  & 1 & 0 & 0 & $\xi_2$ \\
        $\text{U}(1)_3$ & 0 & 0 & $-1$  & $-1$ &1 &1 & $\xi_3$
    \end{tabular}
    \caption{GLSM data for the Abelian Markov quiver.}
    \label{GLSMMarkov}
\end{table}

The three $D$-term equations associated to each U(1) are
\EQ{
    |\phi_1|^2 + |\phi_2|^2 - |\phi_5|^2 - |\phi_6|^2 & = \xi_1\, , \\
    |\phi_3|^2 + |\phi_4|^2 - |\phi_1|^2 - |\phi_2|^2 & = \xi_2\, , \\
    |\phi_5|^2 + |\phi_6|^2 - |\phi_3|^2 - |\phi_4|^2 & = \xi_3\, .
\label{MarkovD}    
}
Again, there is one redundant $D$-term, which gives a consistency constraint $\xi_1+\xi_2+\xi_3=0$. In a certain phase of interest, one can eliminate one of them as was done in \cite{Feng:2000mi} and this elimination corresponds to a decoupling of $\text{U}(1)_+\subset \text{U}(1)_1\times \text{U}(1)_2 \times \text{U}(1)_3$ as described in the previous section. 

\begin{figure}[!h]
    \centering
    \includegraphics[scale=0.5]{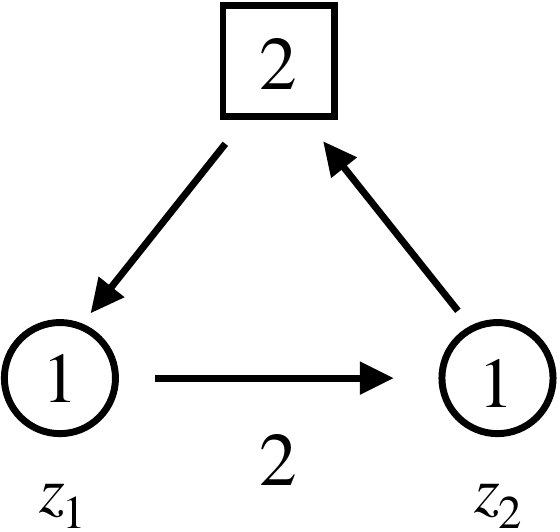}
    \caption{The framed version of the Abelian Markov quiver.}
    \label{FramedMarkov}
\end{figure}

Let us examine its phase structure more closely. When $\xi_1, \xi_2 \gg 0$, the $D$-term constraints tell us the following:
\begin{itemize}
    \item $\xi_2 \gg 0$ requires $\phi_3$ and $\phi_4$ cannot all be zero which gives the base $\mathbb{CP}^1$. Now, the $\phi_1$ and $\phi_2$ describe the fiber directions. Namely, the second D-term equation defines the total space of $\mathcal{O}(-1)\oplus\mathcal{O}(-1)$ over $\mathbb{CP}^1$, and is the same equation of the resolved conifold.
    \item $\xi_1 \gg 0$ further requires $\phi_1$ and $\phi_2$ cannot all vanish, therefore gauging $\text{U}(1)_1$ will give a projectivization of the fibers obtained from $\xi_2\gg 0$, namely $\mathbb{P}\left[ \mathcal{O}(-1)\oplus \mathcal{O}(-1) \right]$. Due to the $\phi_5$ and $\phi_6$, there is another fiber growing on top of the $\mathbb{P}\left[\mathcal{O}(-1)\oplus \mathcal{O}(-1) \right]$.
\end{itemize}
In addition, one can check the above discussion is all consistent with the $D$-term constraint from $\text{U}(1)_3$. Putting everything together, we conclude that this model in the phase $\xi_1, \xi_2 \gg 0$ engineers the following geometry:
\eq{
\left[\mathcal{O}(-1)\oplus\mathcal{O}(-1)\right]_{5,6} \to \mathbb{P}[\mathcal{O}(-1)\oplus\mathcal{O}(-1)]_{1,2} \to   \mathbb{CP}^{1}_{3,4}\,,
}
where we have used the subscripts to distinguish the coordinates. In other words, the redundancy of a $D$-term can effectively freeze one of the gauge nodes into a flavor node, as shown in Fig. \ref{FramedMarkov}.

Let us now look at another phase $\xi_1 \gg 0$ and $\xi_2 \ll 0$. The $D$-term associated with $\xi_1$ determines ${\rm Tot}\left[\mathcal{O}(-1)\oplus \mathcal{O}(-1) \rightarrow \mathbb{CP}^1 \right]$, while the $D$-term associated with $\xi_2$ also gives a geometry described as ${\rm Tot}\left[\mathcal{O}(-1)\oplus \mathcal{O}(-1) \rightarrow \mathbb{CP}^1 \right]$. Note that these two base $\mathbb{CP}^1$'s are the same ones described by $\phi_1$ and $\phi_2$. The $D$-term associated with $\xi_3$ will play an important r\^ole now. Since $\xi_1 +\xi_2 +\xi_3= 0$ and now $\xi_1\gg 0$ and $\xi_2\ll 0$, the sign of $\xi_3$ depends on the sign of $\xi_1 + \xi_2$. 
\begin{itemize}
    \item $\xi_1 + \xi_2 > 0$: in this case $\xi_3 < 0$ and this constrains $\phi_{3,4}$ cannot all be zero, while $\phi_{5,6}$ are along the fibers on the space described by $\phi_{3,4}$. Therefore, now we shall have the same geometry as the phase $\xi_1,\xi_2\gg 0$, but with the coordinates of fibers and the base space exchanged:
\eq{
        \left[\mathcal{O}(-1)\oplus\mathcal{O}(-1)\right]_{5,6} \to \mathbb{P}[\mathcal{O}(-1)\oplus\mathcal{O}(-1)]_{3,4} \to   \mathbb{CP}^{1}_{1,2}\,.
}
    \item $\xi_1 + \xi_2 < 0$: this is similar to the previous case, but again with the fiber and the base coordinates exchanged:
\eq{
        \left[\mathcal{O}(-1)\oplus\mathcal{O}(-1)\right]_{3,4} \to \mathbb{P}[\mathcal{O}(-1)\oplus\mathcal{O}(-1)]_{5,6} \to   \mathbb{CP}^{1}_{1,2}\,.
}
    \item $\xi_1 + \xi_2 = 0$: the third $D$-term equation will identify $\phi_{3,4}$ with $\phi_{5,6}$, which further implies that the first two $D$-term equations become
\EQ{
    |\phi_1|^2 + |\phi_2|^2 - |\phi_3|^2 - |\phi_4|^2 & = \xi_1\,, \\
    -|\phi_1|^2 - |\phi_2|^2 + |\phi_3|^2 + |\phi_4|^2 & = \xi_2\,,
}
    which correspond to a quiver of two gauge nodes connected by four bi-fundamental fields, as shown in Fig. \ref{24quiver}. and in this case another U(1) will further be decoupled. The geometry is
\eq{
        \left[\mathcal{O}(-1)\oplus\mathcal{O}(-1)\right]_{3,4} \oplus  \left[\mathcal{O}(-1)\oplus\mathcal{O}(-1)\right]_{5,6} \to \mathbb{CP}^{1}_{1,2} / \text{Hom} \:\cong\:  \left[\mathcal{O}(-1)\oplus\mathcal{O}(-1)\right] \to \mathbb{CP}^{1} \,,
}
    where the Hom action identifies the fiber directions $\left[\mathcal{O}(-1)\oplus\mathcal{O}(-1)\right]_{3,4}$ with $\left[\mathcal{O}(-1)\oplus\mathcal{O}(-1)\right]_{5,6}$.
    \begin{figure}[!h]
        \centering
        \includegraphics[scale=0.5]{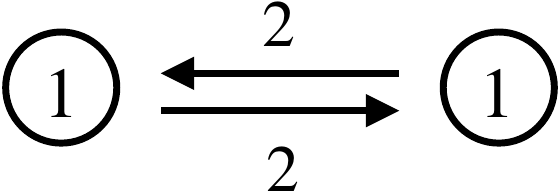}
        \caption{The $k=2$, $n=2$ Abelian necklace quiver at the phase boundary of the Abelian 2-Markov quiver.}
        \label{24quiver}
    \end{figure}
\end{itemize}

\begin{figure}[!h]
\centering
\includegraphics[scale=0.5]{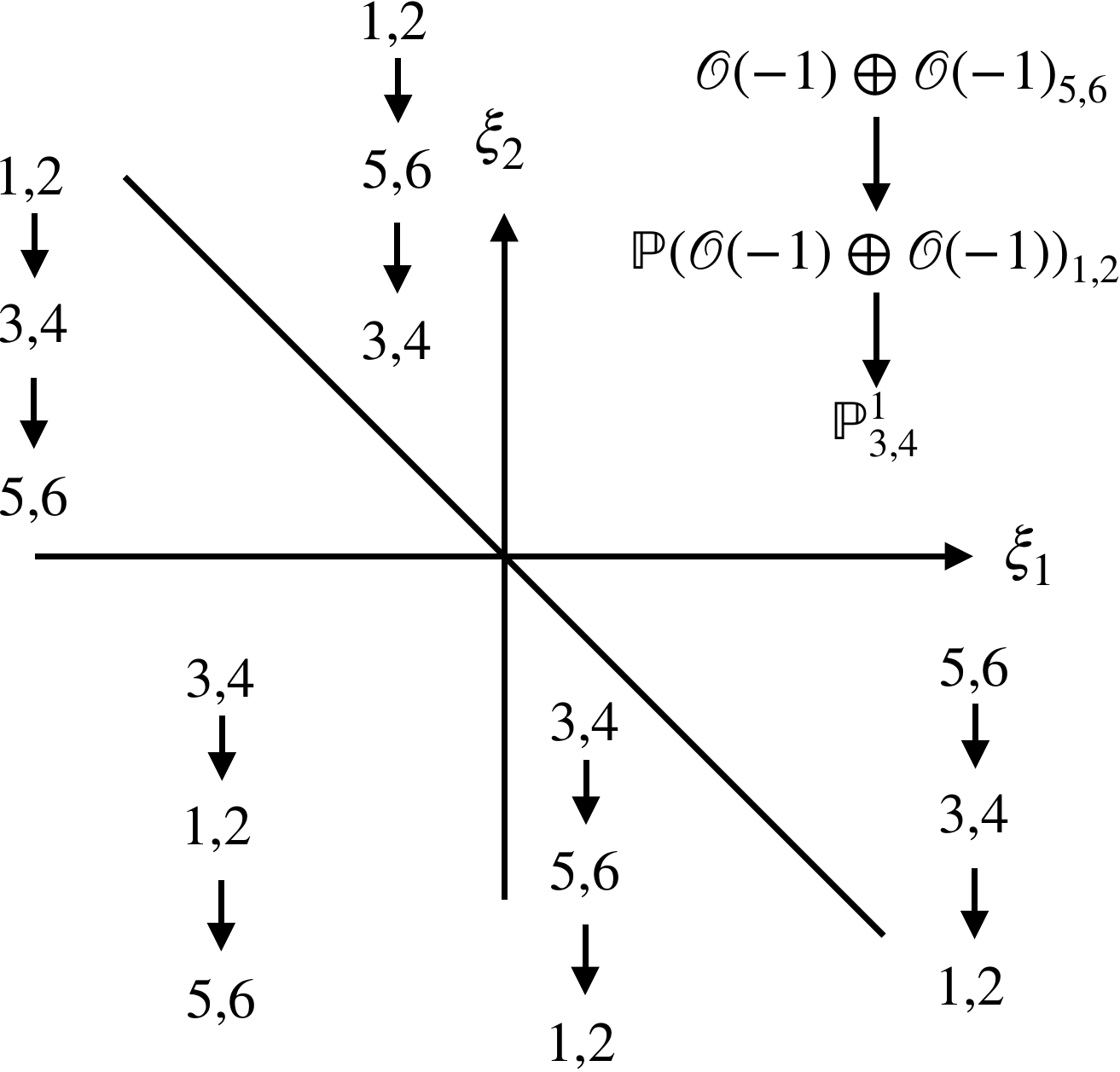}
\caption{The phase diagram of the Abelian Markov quiver without a superpotential.} 
\label{MarkovPhase}
\end{figure}

The geometries in the other phases follow the same argument as above. The whole picture is illustrated in Fig. \ref{MarkovPhase}. Depending on the relative signs, the phase space divides into $3!=6$ chambers, and are related by flop transitions. For example, start with the $\xi_1, \xi_2 > 0$ chamber and cross into the $-\xi_1 < \xi_2 < 0$ chamber, we see that the $(3,4)$ base and the $(1,2)$ fiber are exchanged. If we then cross into the $\xi_2 < - \xi_1 < 0$ chamber, then the $(3,4)$ base exchanges with the $(5,6)$ fiber. Proceeding this way, we permute the base and the fiber, generating all $3!=6$ chambers of the phase space. Curiously, every chamber realizes a geometric phase.

\begin{figure}[!h]
\centering
\includegraphics[scale=0.5]{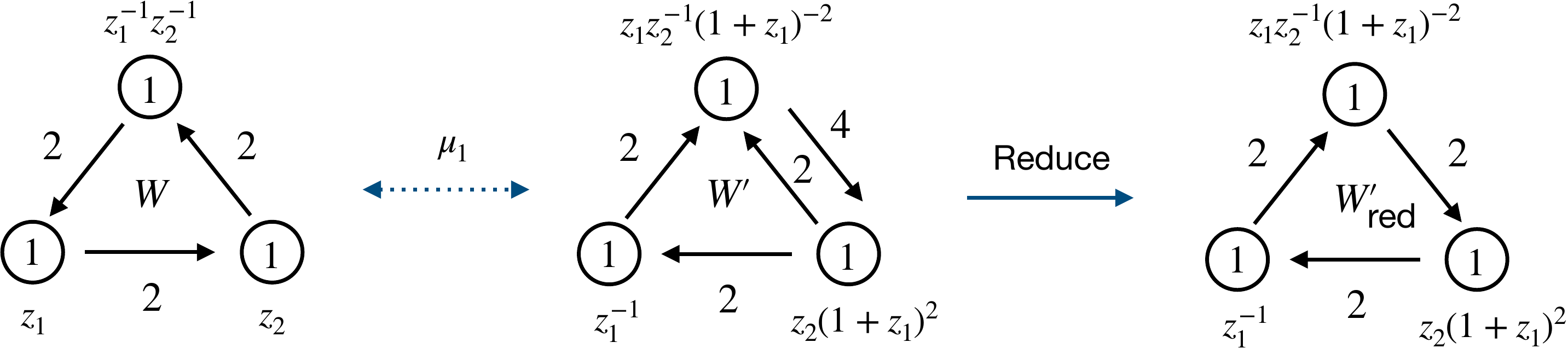}
\caption{The Abelian Markov quiver with a non-degenerate superpotential $W= \phi_5 \phi_3 \phi_1 + \phi_6 \phi_4 \phi_2$ mutates to the same quiver with different FI parameters.}
\label{MarkovMutate}
\end{figure}

A vanishing superpotential, however, would not be compatible with the duality. A mutation, say on the first node, would generate a cubic superpotential of the form
\eq{
W' = M_{15}\, \phi'_5 \phi'_1 +  M_{16}\, \phi'_6 \phi'_1 + M_{25}\, \phi'_5 \phi'_2 +  M_{26}\, \phi'_6 \phi'_2 \,,
}
where $M_{ij}$ are the four new arrows shown in Fig. \ref{MarkovMutate}. The lack of quadratic terms will not allow us to reduce the 2-cycles. A generic superpotential of the form $W = \sum_{\alpha, \beta, \gamma} c_{\alpha \beta \gamma}\, \phi^{5,6}_\gamma \phi^{3,4}_\beta \phi^{1,2}_\alpha$ with $c_{\alpha \beta \gamma} \ne 0$ would also not be compatible with the duality because integrating out the quadratic terms will eliminate all of the terms, leading to a vanishing dual superpotential.

Consider instead the non-degenerate superpotential \cite{Derksen1}
\eq{
W= \phi_5 \phi_3 \phi_1 + \phi_6 \phi_4 \phi_2 \,
.
\label{sup}
}
Under the duality, the combinations $\phi_1 \phi_5$ and $\phi_2 \phi_6$ are replaced by $M_{15}$ and $M_{26}$, respectively.
\eq{
W' = M_{15}\, \phi_3 + M_{26}\, \phi_4 + M_{15}\, \phi'_5 \phi'_1 +  M_{16}\, \phi'_6 \phi'_1 + M_{25}\, \phi'_5 \phi'_2 +  M_{26}\, \phi'_6 \phi'_2 \,.
}
Integrating out the quadratic terms, we obtain a reduced superpotential
\eq{
W'_\text{red}= M_{16}\, \phi'_6 \phi'_1 + M_{25}\, \phi'_5 \phi'_2  \,.
}
Note that the dual superpotential has the same form as \eqref{sup}. Thus one can generate an infinite class of dual geometries without generating a 2-cycle.

The $F$-term equations for the superpotential imply that at least a pair of fields from $\{\phi_1, \phi_3, \phi_5\}$ and at least a pair of fields from $\{\phi_2, \phi_4, \phi_6\}$ will vanish. Depending on the phase of $\xi_1$ and $\xi_2$, the locus in the toric variety will be either a $\mathbb{CP}^1$ or a point. For example, in the $\xi_1, \xi_2 \gg 0$ phase, either $\{\phi_1, \phi_4\}$ or $\{\phi_2, \phi_3\}$ will not vanish. Consider $\phi_1, \phi_4 \ne 0$. The $D$-term equations reduce to
\EQ{
    |\phi_1|^2 & = \xi_1\, , \\
    |\phi_4|^2 - |\phi_1|^2 & = \xi_2\,, \\
 - |\phi_4|^2 & = \xi_3\,.
}
This is just a point after dividing by the gauge group. The case $\phi_2, \phi_3 \ne 0$ is similar.

\begin{figure}[!h]
\centering
\includegraphics[scale=0.5]{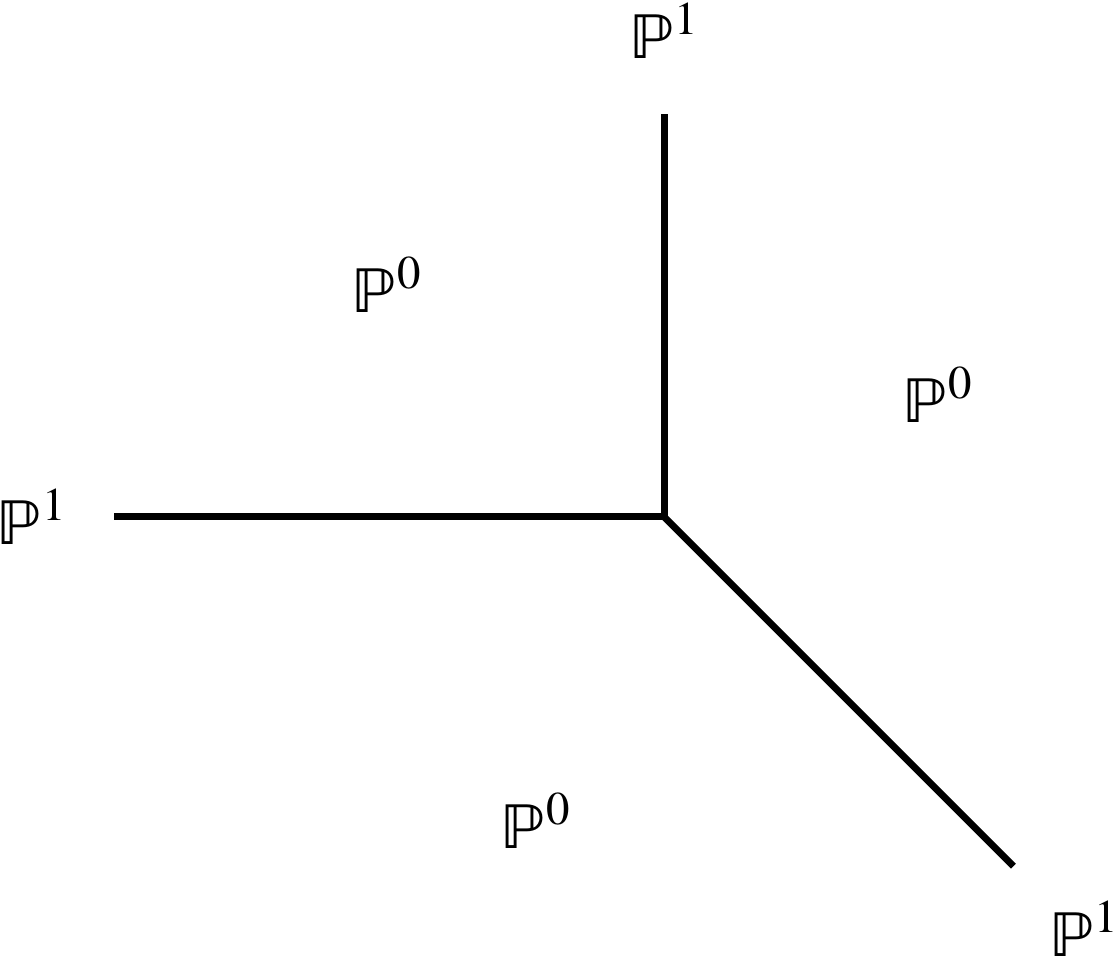}
\caption{The phase diagram of the Abelian Markov quiver with a non-degenerate superpotential $W= \phi_5 \phi_3 \phi_1 + \phi_6 \phi_4 \phi_2$.}
\label{MarkovPhaseW}
\end{figure}

At the phase boundary $\xi_1 + \xi_2 = 0$ where $\xi_1\ll 0$ and $\xi_2 \gg 0$, if either $\{\phi_3, \phi_6\}$ or $\{\phi_4, \phi_5\}$ do not vanish, then we get a point again. Now consider the boundary $\xi_1 + \xi_2 = 0$ where $\xi_1\gg 0$ and $\xi_2\ll 0$. One can only take $\phi_3 = \phi_4 = \phi_5 = \phi_6 = 0$. The $D$-term equations \eqref{MarkovD} then reduce to just one equation
\eq{
- |\phi_1|^2 - |\phi_2|^2  = \xi_2 \,.
}
This is the base $\mathbb{CP}^1$ when $\xi_2 < 0$. The phase structure is shown in Fig. \ref{MarkovPhaseW}. It is consistent with the phases of a general triangle quiver with a generic superpotential \cite[Fig. 3]{Hori:2014tda}, although we emphasize that a generic superpotential would not be compatible with the duality.

On each node, one can perform a duality transformation. The quiver will stay invariant up to an overall $\mathbb{Z}_2$, but the K\"ahler parameters will transform non-trivially as dual cluster variables. This implies that these points in the K\"ahler moduli space should be identified, and the moduli space is tessellated by such equivalence relations.

\paragraph{The $n$-Markov quiver}
This can be generalized to the Abelian $n$-Markov quiver. Note that while the 2-Markov quiver can be identified with the triangulation of the once-punctured torus, the $n$-Markov quiver is not known to arise from a Riemann surface. Nevertheless, we can consider a non-degenerate potential of the form
\eq{
W = \sum_{\alpha=1}^n \phi^{3,1}_\alpha \phi^{2,3}_\alpha \phi^{1,2}_\alpha \, 
\label{WMarkov}
}
After a mutation on the first node, this generates a superpotential according to \eqref{dualW}
\eq{
W' = \sum_{\alpha=1}^n M_{\alpha\alpha}\, \phi^{2,3}_\alpha + \sum_{\alpha,\beta=1}^n M_{\alpha \beta}\, \phi^{1,3}_\beta \phi^{2,1}_\alpha  \,.
}
After integrating out the quadratic terms, the first term is cancelled and we obtain the dual superpotential
\eq{W'_\text{red} = \sum_{\alpha \ne \beta}^n M_{\alpha \beta}\, \phi^{1,3}_\beta \phi^{2,1}_\alpha \,.}

The new terms generated by each successive mutation will cancel the terms in the previous superpotential, eliminating all 2-cycles in the quiver. We thus obtain an infinite number of equivalent descriptions of the hypersurface defined by \eqref{WMarkov} in the non-compact Calabi-Yau space.

\begin{figure}[!h]
\centering    
\begin{tikzpicture}[
  grow cyclic,
  level distance=1cm,
  level/.style={
    level distance/.expanded=\ifnum#1>1 \tikzleveldistance/1.5\else\tikzleveldistance\fi,
    nodes/.expanded={\ifodd#1 fill\else fill=none\fi}
  },
  level 1/.style={sibling angle=120},
  level 2/.style={sibling angle=90},
  nodes={circle,draw,inner sep=+0pt, minimum size=2pt},
  ]
\path[rotate=30]
  node {}
  child foreach \cntI in {1,...,3} {
    node {}
    child foreach \cntII in {1,...,2} { 
      node {}
      child foreach \cntIII in {1,...,2} {
        node {}
        child foreach \cntIV in {1,...,2} {
          node {}
          child foreach \cntV in {1,...,2} {}
        }
      }
    }
  };
  \draw [thick, red, dash pattern=on 0.2cm off 0.2cm]circle(2.5cm);
\end{tikzpicture}
\caption{The infinite mutation sequence of the Markov quiver generates a Bruhat-Tits tree.}
\label{Bruhat-Tits}
\end{figure}
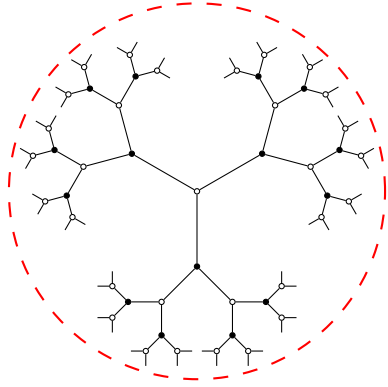

One can perform a sequence of mutations on any of the three nodes, leading to a duality cascade that generates a Bruhat-Tits tree \cite[Fig. 13]{Cachazo:2001sg}, as shown in Fig. \ref{Bruhat-Tits}. A path between any two vertices corresponds to a unique sequence of mutations. The limiting K\"ahler cone in Fig. \ref{LimitRay} corresponds to the boundary of this space. 

\section{The Abelian Necklace Quiver and 2d SQCD}
\label{necklace}

In this section, we will consider the Coulomb branch of the Abelian Markov quiver and its generalization to $k$ nodes. As we shall see, there will be another type of correspondence to a non-Abelian theory: 2d SQCD.

\subsection{Coulomb Branch Analysis}

Consider the Abelian $n$-Markov quiver with massless chiral multiplets. The twisted chiral ring relations \eqref{Baxter} are
\eq{
(\sigma_{i-1}-\sigma_{i})^{n } + (-1)^n z_i (\sigma_{i} - \sigma_{i+1})^{n} = 0 \,,
}
for $i = 1,2,3$. The indices are defined modulo 3. We decouple the overall U(1) by the constraints $z_1 z_2 z_3=1$ and $\sigma_{1,2}+\sigma_{2,3}+\sigma_{3,1} = 0$, where $\sigma_{i,j}:= \sigma_{i} - \sigma_{j}$. Because the solutions $\sigma_i = \sigma_j$ are excluded, we can rewrite the chiral ring relations as
\eq{
\left(\frac{\sigma_{1,2}}{-\sigma_{1,2}-\sigma_{2,3}}\right)^n = e^{t_1}, \qquad \left(\frac{\sigma_{2,3}}{-\sigma_{1,2}-\sigma_{2,3}}\right)^n = e^{t_1 + t_2}\,.
}
Apart from the decoupled U(1) direction, there is no supersymmetric vacuum at generic points in the K\"ahler moduli space. Our key observation is that at the origin $(t_1, t_2) = 0$, the vacuum equations take the same form as SU(3) SQCD with $n$ massless chiral multiplets.

Let us review the vacuum structure of SU($k$) SQCD with $n$ massless chiral multiplets. Since SU($k$) SQCD does not have a U(1) factor, there is no FI parameter. One also cannot turn on a $\theta$ angle because  $\pi_1(\text{SU}(k)) = 0$. The twisted superpotential \eqref{twistedW} is
\eq{
\widetilde W_\text{eff} = -i\pi\sum^k_{a < b} (\sigma_a - \sigma_b) - n \sum_{a=1}^{k}  \sigma_a \left( \log \sigma_a  -\frac{i\pi}{2}  -1 \right) \,,
}
subject to the constraint $\sum_{a=1}^k \sigma_a = 0$. The vacuum equations are 
\eq{
\left(\frac{\sigma_a}{-\sigma_1 - \sigma_2 - \cdots - \sigma_{k-1}}\right)^n = 1\,.
\label{SQCDvacua}
}
for $a=1,2,\ldots, k-1$.
Alternatively, one can regard all $\sigma_a$'s as independent fields and impose the traceless condition by a Lagrange multiplier $\lambda \in \mathbb{C}$. That is, find the extremum of 
\eq{\widetilde W_\text{eff} - \lambda \sum_{a=1}^k \sigma_a  \,,}
with respect to $\sigma_a$ and $\lambda$. Thus the vacuum equations \eqref{SQCDvacua} are equivalent to 
\eq{\sigma_a^n = e^{-\lambda}\,.
\label{Lagrange}
} 
subject to the traceless constraint.

An important feature of massless SQCD is that, for certain choices of $(k,n)$, the vacuum will develop flat directions  \cite{Hori:2006dk}.
When $k=2$, $\widetilde W_\text{eff} = i n \pi \sigma_1$ is effectively a $\theta$-angle term with $\theta = n \pi$, which induces a background electric field. If $n$ is even, then pair creation of charged particles will screen the electric field and $\sigma_1$ is unconstrained. If $n$ is odd, then there is an effective $\theta = \pi$ and the vacuum will be lifted.

Consider $k=3$, whose vacuum equation coincides with that of the Abelian Markov quiver. If $n$ is a multiple of 3, then a one-dimensional non-compact Coulomb branch appears in the direction $(\sigma_{1}, \sigma_{2}, \sigma_{3}) = (1, e^{2\pi i/3}, e^{4\pi i/3})\sigma$. An overall scaling will not change the direction, and solutions related by permutations are identified under the Weyl group. Singularities correspond to $k$ distinct $n$-th roots of unity, modulo overall scaling, that sum to zero. For higher rank, multiple Coulomb branch directions may open up. For $(k,n) = (4,8)$, there are two non-compact directions along $(\sigma_{1}, \sigma_{2}, \sigma_{3},\sigma_{4}) = (1,-1,i,-i)\sigma$ and $(1,-1,e^{i\pi/4}, -e^{i\pi/4})\sigma$. We tabulate the number of non-compact Coulomb branches in Table \ref{qCoulomb}. 

\begin{table}[!h]
\center
\begin{tabular}{c|c|c|c|c|c|c|c|c|c|c|c}
$k \backslash n$ & 2 & 3 & 4 & 5 & 6 & 7 & 8 & 9 & 10 & 11 & 12 \\
\hline
2 & 1 & 0 & 1 & 0 & 1 & 0 & 1 & 0 & 1 & 0 & 1 \\
3 & 0 & 1 & 0 & 0 & 1 & 0 & 0 & 1 & 0 & 0& 1  \\
4 & 0 & 0 & 1 & 0 & 1 & 0 & 2 & 0 & 2 & 0 & 3 \\ 
5 & 0 & 0 & 0 & 1 & 0 & 0 & 0 & 0 &1 & 0 & 1 \\ 
6 & 0 & 0 & 0 &  0 & 1 & 0 & 1& 1 & 2 & 0 & 5 \\
7 & 0 & 0 & 0 & 0 & 0 & 1 & 0 & 0 & 0 & 0& 1  \\
\end{tabular}
\caption{The number of non-compact Coulomb branches of SU($k$) SQCD with $n$ massless fundamental chiral multiplets. } 
\label{qCoulomb}
\end{table}

Theories that admit such singularities are known as irregular theories and theories that do not have such singularities are known as regular theories \cite{Hori:2011pd}. A natural question is: when is a theory regular? A result from number theory states that one can find $k$ distinct $n$-th roots of unity that sum to zero if and only if both $k$ and $n-k$ can be written as a linear combination of prime factors of $n$ with nonnegative integer coefficients \cite{Sivek_2010}. This immediately implies that SU($k$) SQCD is regular precisely when this condition is not satisfied. In particular, SU($k$) SQCD with a prime number $p$ of chiral multiplets is regular for all $k<p$. 

We return to the Abelian Markov quiver and perform a similar analysis.
If $n$ is a multiple of 3, then a pair of one-dimensional non-compact Coulomb branches appear in the directions $(\sigma_{1,2}, \sigma_{2,3}, \sigma_{3,1}) = (1, e^{2\pi i/3}, e^{4\pi i/3})\sigma$ and $(1, e^{4\pi i/3}, e^{2\pi i/3})\sigma$. There is an important caveat that makes the analysis differ from that of the non-Abelian theory.  For non-Abelian gauge groups, Cartan elements related by Weyl symmetry are identified. Such Weyl symmetry is absent in the Abelian quiver. We will find many more solutions related by permutations. 

What about SU($k$) SQCD with $n$ massless chiral multiplets? This naturally leads us to consider an $n$-necklace quiver with $k$ nodes. The Abelian 2-Markov quiver corresponds to $(k,n) = (3,2)$. The vacuum equations read
\eq{
\left(\frac{\sigma_{i,i+1}}{\sigma_{i-1,i}}\right)^n = e^{t_i} \,,
\label{vacuum}
}
for $i=1, 2, \ldots, k$. If we write $\sigma_{i,i+1} = s_i \sigma$, with $s_1 = 1$, then the product \eqref{vacuum} telescopes and we obtain
\eq{
s_m^{\,n} = \prod_{i=2}^m e^{t_i}\,,
\label{tele}
} 
subject to the constraint $\sum_{i=1}^k s_i = 0$. At the origin of the K\"ahler moduli space, this coincides with the vanishing sums of unity condition for SU($k$) SQCD.
Consider $(k, n)=(4,2)$. we find three flat directions along $\vec s = (1,-1,-1,1)$, $(1,-1,1,-1)$ and $(1,1,-1,-1)$. Here we encounter another caveat.  Since the vacuum equations are functions of $\sigma_{i,j}$ instead of $\sigma_i$, a degenerate solution $\sigma_{i,j} = \sigma_{k,l}$ \eqref{tele} is not necessarily a degenerate vacuum of the quiver theory. We only need to exclude solutions where $\sigma_{i,i+1} = 0$, which is guaranteed by the form of the vacuum equations \eqref{vacuum}. For example, the $\vec s = (1,-1,1,-1)$ solution corresponds to $\sigma_1 = \sigma_3, \sigma_2 = \sigma_4$, which is a valid weakly-coupled solution for the necklace quiver. The singularities correspond to $k$ possibly repeating $n$-th roots unity, modulo overall scaling, that sum to zero. When repetitions are allowed,  the vanishing sums condition is satisfied if and only if $k$ can be written as a linear combination of prime factors of $n$ with nonnegative integer coefficients \cite{Lam_2000}. We tabulate the number of non-compact Coulomb branches in Table \ref{qCoulomb1}. 

\begin{table}[!h]
\center
\begin{tabular}{c|c|c|c|c|c|c|c|c|c|c|c}
$k \backslash n$ & 2 & 3 & 4 & 5 & 6 & 7 & 8 & 9 & 10 & 11 & 12\\
\hline
2 & 1 & 0 & 1 & 0 & 1 & 0 & 1 & 0 & 1 & 0 & 1\\
3 & 0 & 2 & 0 & 0 & 2 & 0 & 0 & 2 & 0 & 0 & 2\\
4 & 3  & 0 & 9 & 0 &  15  & 0 & 21& 0 & 27 & 0 & 33\\
5 & 0 &  0  & 0 &  24 & 60 & 0 & 0 & 0 & 24 & 0 & 180 \\
6 &  10 & 30 & 100 &  0 & 340 & 0 & 640 & 270 & 1090 & 0 & 1930 \\
7 & 0 & 0 & 0 & 0 & 1680 & 720 & 0 & 0 & 2520 & 0 & 15540 
\end{tabular}
\caption{The number of non-compact Coulomb branches of the Abelian $n$-necklace quiver with $k$ nodes at the origin of the K\"ahler moduli space.}
\label{qCoulomb1}
\end{table}

Thus the Abelian necklace quiver can be regarded as an Abelianization of 2d SQCD. The non-compact Coulomb branches of SQCD can be found by imposing the Weyl group and eliminating repeated roots. For $n < k$, all solutions to the vacuum equations \eqref{vacuum} are degenerate. They are discarded in SQCD. We stress that it is the difference of Coulomb vacua between neighboring nodes in the Abelian quiver that maps to the Coulomb vacua of SQCD
\eq{
(\sigma_{i,i+1})^\text{Abelian quiver} \quad \longleftrightarrow \quad (\sigma_i)^\text{SQCD} \,.
}

\subsection{Singular Loci in the K\"ahler Moduli Space}
The foregoing analysis is essentially an Abelianization of Hori and Tong's analysis of SU($k$) SQCD. But we now have the additional freedom to tune the FI parameters. The K\"ahler moduli space is $k$-dimensional.  There are $k-1$ independent K\"ahler coordinates $t_i$, $i = 1,2,\ldots, k-1$, which determine the $k$-th coordinate by $\sum_{i=1}^k t_i = 0$ modulo $2\pi i$.

When $k=2$, we have seen that the origin is singular when $n$ is even. We find that there is another singularity at $\theta_1 = \pi$, where the theory is regular at even $n$, but is singular at odd $n$.

When $k=3$, the point $\theta_i = \pi$ for $i=1,2$ is regular for any $n$ so is a smooth point on the moduli space. This is in distinction with SQCD at $\theta=0$ where there is at least one $n$ that is irregular. We find additional singularities when a discrete $\theta$ angle is turned on at $\theta_i=2\pi i/3$ and $\theta_i=4\pi i/3$.
We tabulate the number of non-compact Coulomb branches at $\theta_i = \pi$ in Table \ref{qCoulomb2}. 

\begin{table}[!h]
\center
\begin{tabular}{c|c|c|c|c|c|c|c|c|c|c|c|c}
$k \backslash n$ & 1 & 2 & 3 & 4 & 5 & 6 & 7 & 8 & 9 & 10 & 11 & 12\\
\hline
2 & 1 & 0 & 1 & 0 & 1 & 0 & 1 & 0 & 1 & 0 & 1 & 0\\
3 & 0 & 0 & 0 & 0 & 0 & 0 & 0 & 0 & 0 & 0 & 0 & 0\\
4 & 1 & 2  & 5 & 4 & 9 &  6 & 13 & 8 & 17& 10 & 21 & 12\\
5 & 0 & 0 & 0  & 0 & 0 & 12 & 0 &  0 & 0 & 0 & 0 & 24 \\
6 & 1 & 0 & 31 & 0 & 109 & 24 & 235 & 0 & 433 & 0 & 631  & 48  \\
7 & 0 & 0 & 0 & 0 & 0 & 180 & 0 & 0 & 0 & 0 & 0 & 792 
\end{tabular}
\caption{The number of non-compact Coulomb branches of the Abelian $n$-necklace quiver with $k$ nodes at another singularity of the K\"ahler moduli space, $t_i = i \pi$ for $i=1, 2, \ldots,k-1$.}
\label{qCoulomb2}
\end{table}

\begin{figure}[!h]
\centering
\includegraphics[scale=0.5]{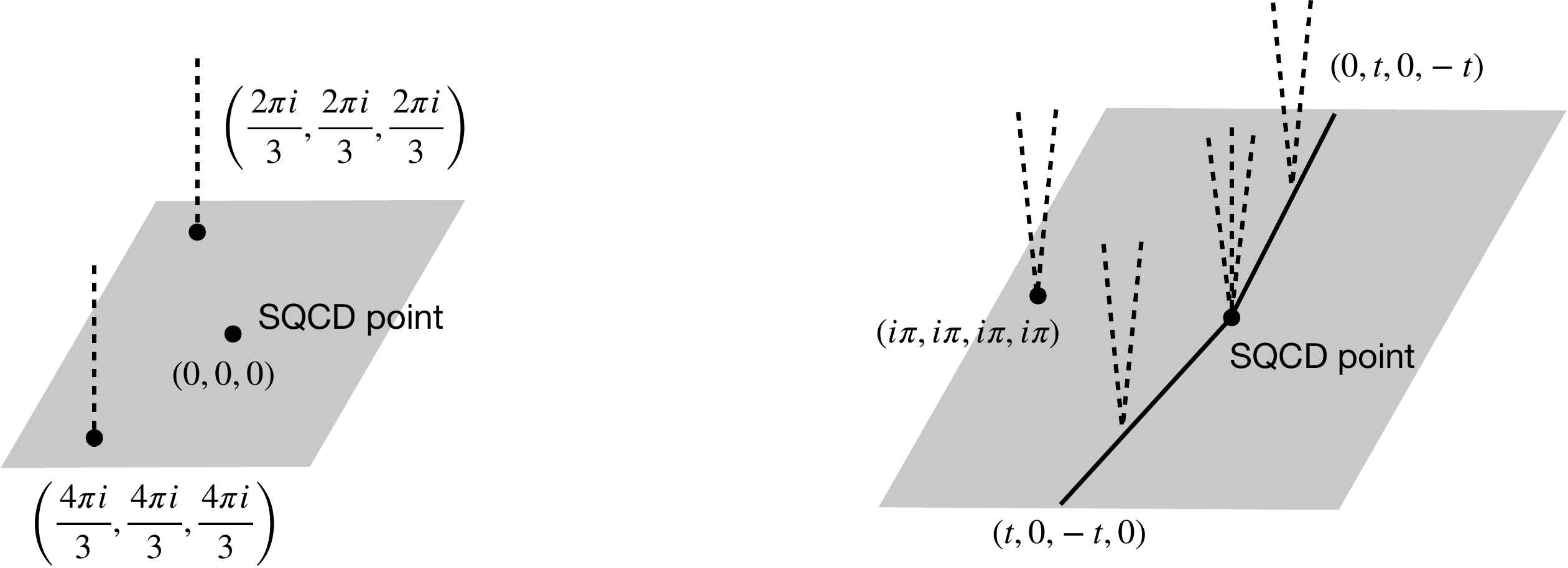}
\caption{The K\"ahler moduli spaces of the necklace quiver for $(k,n)=(3,4)$ and $(k,n)=(4,2)$. As we tune the complexified FI parameters, both isolated singular points and continuous singular loci, denoted by solid lines, may appear. At each point on the singular loci, multiple one-dimensional non-compact Coulomb branches emerge, as shown in dashed lines.}
\label{Kmoduli}
\end{figure}

A new phenomenon arises when $k > 3$. There is a continuous family of solutions that support non-compact Coulomb branches as we tune the FI parameters. In particular, the degenerate solutions that were previously excluded from the SQCD analysis become non-degenerate. Consider $k=4$. Let us move in the K\"ahler moduli space along the direction $\vec t = (t, 0, -t, 0)$ or along the direction $\vec t = (0, t, 0, -t)$. For $(k,n) = (4,2)$, we find a pair of non-compact Coulomb branches emanating in the $\vec s = (1,-1,\pm \frac{1}{\sqrt{z}}, \mp\frac{1}{\sqrt{z}})$ directions on the $\vec t = (t, 0, -t, 0)$ loci, and another pair in the $\vec s = (1,\pm \sqrt{z}, \mp \sqrt{z}, -1)$ directions on the $\vec t = (0, t, 0, -t)$ loci. The singular loci are shown in Fig. \ref{Kmoduli}. 

Thus in addition to isolated singular points, we find one-dimensional singular loci. 
How can we interpret the appearance of such singular loci? Recall that each Abelian node can be thought of as a Cartan element of SU($k)$ and turning on FI parameters for the Abelian quiver corresponds to turning on an ``effective FI parameter" $t_2 + t_3 + \cdots + t_m$ for the $m$-th Cartan element \eqref{tele}. For example, on the $\vec t = (t, 0, -t, 0)$ loci, the vacuum equations are
\eq{
s^n_1 = 1, \qquad s^n_2 = 1, \qquad s^n_3 = e^{-t}, \qquad s^n_4 = e^{-t} \,.
}
Thus the vacuum equations split into a pair of equations for SU($2$) SQCD \eqref{Lagrange}, with the effective FI parameters playing the r\^ole of the Lagrange multipliers. The first SU(2) supports a non-compact Coulomb branch along $(1,-1)$, while the second SU(2) supports a non-compact Coulomb branch along $(\frac{1}{\sqrt{z}},-\frac{1}{\sqrt{z}})$. We note that instead of being a Lagrange multiplier that is to be solved for the critical point, $t$ is a free variable that parametrizes the singular loci. 

A singular locus can be interpreted as a Lagrange multiplier for each unbroken subgroup. This allows us to find higher-dimensional singular loci in higher-rank theories, e.g. $\vec t = (t_1, 0, -t_1, t_2, 0, -t_2)$ for $(k,n)  = (6,2)$. Thus we find singular loci of mixed dimensions in the K\"ahler moduli space. We leave a more systematic study of the singularity structure to future work.

\section*{Acknowledgments}
We thank Yingchun Zhang for raising a question on unframed quivers that led to this note. We thank Jin Chen, Wei Cui, Babak Haghighat, Taro Kimura, Sungjay Lee, Mauricio Romo, Eric Sharpe, Fengjun Xu, Junya Yagi, and especially Prajna Lin for discussions. We thank Hongfei Shu and Ruidong Zhu for collaboration on a related project. PZ is supported in part by a China Postdoctoral Science Foundation Special Fund, grant number Y9Y2231.

\bibliographystyle{JHEP}
\bibliography{ref}

\appendix

\end{document}